\pdfoutput=1

\documentclass[11pt,twoside,a4paper,cmspaper,final,collab]{cms-tdr}
\def\svnVersion{486741P}\def\cmsCernNoTag{CERN-EP-2018-195}\def\cmsCernDate{\today}\def\cmsMessage{Published in the Journal of High Energy Physics as \href{http://dx.doi.org/10.1007/JHEP01(2019)040}{\doi{10.1007/JHEP01(2019)040.}}}

\begin{document}\cmsNoteHeader{B2G-17-019}

\hyphenation{had-ron-i-za-tion}
\hyphenation{cal-or-i-me-ter}
\hyphenation{de-vices}
\RCS$HeadURL: svn+ssh://svn.cern.ch/reps/tdr2/papers/B2G-17-019/trunk/B2G-17-019.tex $
\RCS$Id: B2G-17-019.tex 486694 2019-01-21 06:27:21Z devdatta $

\providecommand{\cmsLeft}{left\xspace}
\providecommand{\cmsRight}{right\xspace}

\providecommand{\CL}{CL\xspace}
\providecommand{\cmsTable}[1]{\resizebox{\textwidth}{!}{#1}}
\newcommand{\Hbb}{\ensuremath{\PH\to\bbbar}\xspace}
\newcommand{\ttjets}{\ttbar{}+jets\xspace}
\newcommand{\intLumi}{35.9\fbinv}
\newcommand{\Hbbt}{double-\PQb\,tagger\xspace}
\newcommand{\mjj}{\ensuremath{m_{\text{jj}}}\xspace}
\newcommand{\mjjj}{\ensuremath{m_{\text{Jjj}}}\xspace}
\newcommand{\mjjjs}{\ensuremath{m_{\text{Jjj,red}}}\xspace}
\newcommand{\mH}{\ensuremath{m_{\PH}}\xspace}
\newcommand{\mHH}{\ensuremath{m_{\PH\PH}}\xspace}
\newcommand{\nsub}{\ensuremath{\tau_{21}}\xspace}
\newcommand{\LambdaR}{\ensuremath{\Lambda_{\text{R}}}\xspace}
\newcommand{\PX}{\ensuremath{\cmsSymbolFace{X}}\xspace}
\newcommand{\mx}{\ensuremath{m_{\PX}}\xspace}
\providecommand{\NA}{\ensuremath{\text{---}}}
\newcolumntype{P}[1]{>{\centering\arraybackslash}m{#1}}

\cmsNoteHeader{B2G-17-019}

\title{Search for production of Higgs boson pairs in the four $\cPqb$ quark final state using large-area jets in proton-proton collisions at $\sqrt{s}=13\TeV$}

\date{\today}

\abstract{A search is presented for pair production of the standard model Higgs boson using data from proton-proton collisions at a centre-of-mass energy of 13\TeV, collected by the CMS experiment at the CERN LHC in 2016, and corresponding to an integrated luminosity of 35.9\fbinv. The final state consists of two $\cPqb$ quark-antiquark pairs. The search is conducted in the region of phase space where one pair is highly Lorentz-boosted and is reconstructed as a single large-area jet, and the other pair is resolved and is reconstructed using two $\cPqb$-tagged jets. The results are obtained by combining this analysis with another from CMS looking for events with two large jets. Limits are set on the product of the cross sections and branching fractions for narrow bulk gravitons and radions in warped extra-dimensional models having a mass in the range 750--3000\GeV. The resulting observed and expected upper limits on the non-resonant Higgs boson pair production cross section correspond to 179 and 114 times the standard model value, respectively, at 95\% confidence level. The existence of anomalous Higgs boson couplings is also investigated and limits are set on the non-resonant Higgs boson pair production cross sections for representative coupling values.}

\hypersetup{
pdfauthor={CMS Collaboration},
pdftitle={Search for production of Higgs boson pairs in the four b quark final state using large-area jets in proton-proton collisions at sqrt(s)=13 TeV},
pdfsubject={CMS},
pdfkeywords={CMS, physics, Higgs, beyond standard model, jets}}

\maketitle

\section{Introduction\label{sec:Introduction}}

In the standard model (SM), Higgs boson (\PH)~\cite{HiggsDiscoveryAtlas,HiggsDiscoveryCMS,CMSHiggsLongPaper} pair-production can occur through several subprocesses and is sensitive to the Higgs boson self-coupling.
In proton-proton (\Pp\Pp) collisions at the CERN LHC, the SM $\PH\PH$ production cross section is mainly due to the
gluon-gluon fusion subprocess, which proceeds via an internal fermion loop dominated by the top quark, \cPqt.
At a centre-of-mass energy of 13\TeV, this cross section is $33.5^{+2.5}_{-2.8}$\unit{fb}~\cite{deFlorian:2016spz,deFlorian:2013jea,Baglio:2012np}, which is too small to be observable using the current data.
However, many beyond the standard model (BSM) theories predict higher rates of Higgs boson pair production.
The rate could be increased through the production of a massive BSM resonance \PX, which subsequently decays to a Higgs boson pair ($\PX\to\PH\PH$)~\cite{Randall:1999ee}, a process that could be observable at the LHC.
If the resonance mass \mx is too large for \PX to be directly produced in $\Pp\Pp$ interactions, the particle could  manifest itself through off-shell effects, leading to anomalous couplings of the $\PH$ boson to the SM particles, including the $\PH\PH$ self-interaction~\cite{Grober:2010yv}.
Thus, BSM effects may modify the $\PH\PH$ differential and integral production cross sections, making this process observable with current data.

Models with a warped extra dimension (WED), as proposed by Randall and Sundrum~\cite{Randall:1999ee}, are among those BSM scenarios that predict the existence of resonances with large couplings to the SM Higgs boson, such as the spin-0 radion~\cite{Goldberger:1999uk,DeWolfe:1999cp,Csaki:1999mp} and the spin-2 first Kaluza--Klein (KK) excitation of the graviton~\cite{Davoudiasl:1999jd,Csaki:2000zn, Agashe:2007zd}.
The WED models postulate an additional spatial dimension $l$ compactified between two four-dimensional hypersurfaces known as the branes, with the region between, the bulk, warped by an exponential metric $\kappa l$, where  $\kappa$ is the warp factor~\cite{Giudice:2000av}.
A value of $\kappa l\!\!\sim\!\!35$ fixes the mass hierarchy between the Planck scale \Mpl and the electroweak scale~\cite{Randall:1999ee}.
One of the parameters of the model is $\kappa/\overline{\Mpl}$, where $\overline{\Mpl} \equiv \Mpl /\sqrt{8\pi}$.
The ultraviolet cutoff scale of the model $\LambdaR \equiv \sqrt{6} \re^{-\kappa l} \overline{\Mpl}$~\cite{Goldberger:1999uk} is another parameter, and is expected to be near the TeV scale.

In the absence of new resonances coupling to the Higgs boson, the gluon fusion Higgs boson pair production subprocess can still be enhanced by BSM contributions to the coupling parameters of the Higgs boson and the SM fields~\cite{Dawson:2015oha}. The SM production rate of $\PH\PH$ through gluon fusion is determined by the Yukawa coupling of the Higgs boson to the top quark $y^{\text{SM}}_{\cPqt}$ and the Higgs boson self-coupling $\lambda^{\text{SM}}_{\PH\PH}=\mH^2/2v^2$. Here, $\mH=125\GeV$ is the Higgs boson mass~\cite{Aad:2015zhl,CMS2016HZZ4l} and $v=246\GeV$ is the vacuum expectation value of the Higgs field.
Deviations from the SM values of these two coupling parameters can be expressed as $\kappa_{\lambda} \equiv \lambda_{\PH\PH}/\lambda^{\text{SM}}_{\PH\PH}$ and $\kappa_{\cPqt} \equiv y_{\cPqt}/y^{\text{SM}}_{\cPqt}$, respectively.
Depending on the BSM scenario, other couplings not present in the SM may also exist and can be described by dimension-6 operators in the framework of an effective field theory by the Lagrangian~\cite{Giudice:2007fh}:
\begin{linenomath}
\begin{equation*}
\begin{aligned}
\mathcal{L}_\PH =
& \frac{1}{2} \partial_{\mu}\, \PH \partial^{\mu} \PH - \frac{1}{2} \mH^2 \PH^2 -
  {\kappa_{\lambda}}\,  \lambda^{\text{SM}}_{\PH\PH} v\, \PH^3
- \frac{ m_\cPqt}{v}(v+   {\kappa_\cPqt} \,   \PH  +  \frac{c_{2}}{v}   \, \PH\PH ) \,( \cPaqt_{\text{L}}\cPqt_{\text{R}} + \text{h.c.}) \\
& + \frac{1}{4} \frac{\alpha_s}{3 \pi v} (   c_\cPg \, \PH -  \frac{c_{2\Pg}}{2 v} \, \PH\PH ) \,  G^{\mu \nu}G_{\mu\nu}\,.
\label{eq:lag}
\end{aligned}
\end{equation*}
\end{linenomath}
The anomalous couplings and the corresponding parameters in this Lagrangian are: the contact interaction between a pair of Higgs bosons and a pair of top quarks ($c_{2}$), the interaction between the Higgs boson and the gluon ($c_\Pg$), and the interaction between a pair of Higgs bosons and a pair of gluons ($c_{2\Pg}$).
The couplings with CP-violation and the interactions of the Higgs boson with light SM and BSM particles are not considered.
The Lagrangian models the effects of BSM scenarios with a scale that is beyond the direct LHC reach.
This five-parameter space of BSM Higgs couplings has constraints from measurements of single Higgs boson production and other theoretical considerations~\cite{Khachatryan:2014jba,Aad:2015gba}.

Searches for $\PH\PH$ production have been performed by the ATLAS \cite{Aad:2014yja,Aad:2015uka,Aad:2015xja,Aaboud:2016xco,ATLAS_HH4b_2016_13TeV,ATLAS_bbtautau_2016_13TeV,ATLAS_HHWWgamgam_2016_13TeV,ATLAS_HHbbgamgam_2016_13TeV} and CMS \cite{Khachatryan:2014jya,Sirunyan:2017tqo,Khachatryan:2015year,Khachatryan:2015tha,Khachatryan:2016sey,Khachatryan:2016cfa,CMSHHbbtatau13TeV2016data,Sirunyan:2017guj,CMS-HIG-17-008} Collaborations using the LHC $\Pp\Pp$ collision data at $\sqrt{s}=8$ and 13\TeV.
A search targeting the high \mx range for a KK bulk graviton or a radion decaying to $\PH\PH$, in the $\bbbar\bbbar$ final state, was published by the CMS Collaboration~\cite{CMS-B2G-16-026}, in which two large-area jets are used to reconstruct the highly Lorentz-boosted Higgs bosons (``fully-merged'' event topology).
A similar search, focusing on a lower range of \mx, was also performed by CMS~\cite{CMS-HIG-17-009}, using events with four separate $\cPqb$ quark jets. The configuration of a Higgs boson candidate as one large-area jet or as two separate smaller jets is dependent on the momentum of the Higgs boson~\cite{Gouzevitch:2013qca}.

In this paper, we improve upon the CMS search for high mass resonance ($750 \le \mx \le 3000\GeV$) decaying to $\PH\PH \to \bbbar\bbbar$~\cite{CMS-B2G-16-026} by using ``semi-resolved'' events, \ie those containing exactly one highly Lorentz-boosted Higgs boson while the other Higgs boson is required to have a lower boost.
The data set corresponds to an integrated luminosity of \intLumi from $\Pp\Pp$ collisions at 13\TeV.
The more boosted Higgs boson is reconstructed using a large-area jet and the other is reconstructed from two separate $\cPqb$ quark jets.
The inclusion of the semi-resolved events leads to a significant improvement in the search sensitivity for resonances with $750 \le \mx \le 2000\GeV$.
With the addition of the semi-resolved events, a signal from the non-resonant production of $\PH\PH$ is also accessible using boosted topologies, since such production typically results in an $\PH\PH$ invariant mass that is lower than that of a postulated resonance signal.
For full sensitivity, the results are obtained using a statistical combination of the semi-resolved events with the fully-merged events selected using the criteria in Ref.~\cite{CMS-B2G-16-026}.
In addition to improving the search for $\PX \to \PH\PH$, strong constraints are thus obtained for several regions in the $\PH$ boson anomalous coupling parameter space, defined by Eq.~(\ref{eq:lag}).

\section{The CMS detector and event reconstruction\label{sec:CMSDetector}}

The CMS detector with its coordinate system and the relevant kinematic
variables is described in Ref.~\cite{CMSDetector}. The central
feature of the CMS apparatus is a superconducting solenoid of
6\unit{m} internal diameter, providing a magnetic field of
3.8\unit{T}. Within the field volume are silicon pixel and strip
trackers, a lead tungstate crystal electromagnetic calorimeter (ECAL),
and a brass and scintillator hadron calorimeter (HCAL), each composed
of a barrel and two endcap sections. The tracker covers a
pseudorapidity $\eta$ range from $-2.5$ to 2.5 with the ECAL and the
HCAL extending up to $\abs{\eta}=3$. Forward calorimeters in the region
up to $\abs{\eta}=5$ provide good hermeticity to the detector.
Muons are detected in gas-ionization chambers embedded
in the steel flux-return yoke outside the solenoid, covering a region of
$\abs{\eta}<2.4$.

Events of interest are selected using a two-tiered trigger system~\cite{CMSTrigger}. The first level (L1), composed of custom hardware processors, uses information from the calorimeters and muon detectors to select events at a rate of around 100\unit{kHz}. The second level, known as the high-level trigger (HLT), consists of a farm of processors running a version of the full event reconstruction software optimized for fast processing, and reduces the event rate to around 1\unit{kHz} before data storage. Events used in this analysis are selected at the trigger level based on the presence of jets in the detector. The level-1 trigger algorithms reconstruct jets from energy deposits in the calorimeters. The particle-flow (PF) algorithm~\cite{CMS-PF-GED}, aims to reconstruct and identify each individual particle in an event. The physics objects reconstructed include jets (clustered with a different algorithm), electrons, muons, photons, and also the missing-\pt vector.

Multiple $\Pp\Pp$ collisions may occur in the same or adjacent LHC bunch crossings
(pileup) and contribute to the overall event activity in the
detector.
The reconstructed vertex with the largest value of summed physics-object $\pt^{2}$ is taken to be the primary $\Pp\Pp$ interaction vertex. The physics objects are the jets, clustered using the jet finding algorithm~\cite{antiKtAlgorithm,FastJet} with the tracks assigned tot he vertex as inputs, and the associated missing transverse momentum, taken as the negative vector sum of the \pt of those jets.
The other interaction vertices are designated as pileup vertices.

The energy of each electron is determined from a combination of the electron momentum at the primary
interaction vertex as determined by the tracker, the energy of the corresponding ECAL cluster, and the
energy sum of all bremsstrahlung photons spatially compatible with originating from the electron track. The energy of each muon is obtained from the curvature of the corresponding track.
The energy of each charged hadron is determined from a combination of its
momentum measured in the tracker and the matching ECAL and HCAL energy deposits, corrected for
zero-suppression effects and for the response function of the calorimeters to hadronic
showers. Finally, the energy of each neutral hadron is obtained from the corresponding corrected ECAL and HCAL energies.

Particles reconstructed by the PF algorithm are clustered into jets with the anti-\kt algorithm~\cite{antiKtAlgorithm, FastJet}, using a distance parameter of 0.8 (AK8 jets) or 0.4 (AK4 jets). The jet transverse momentum is determined as the vector sum \pt of all clustered particles. To mitigate the effect of pileup on the AK4 jet momentum, tracks identified as originating from pileup vertices are discarded in the clustering, and an offset correction~\cite{jetarea_fastjet_pu,Khachatryan:2016kdb} is applied for remaining contributions from neutral particles. Jet energy corrections are derived from simulation to bring the measured response of the jets to that of particle level jets on average. In situ measurements of the momentum balance in events containing either a pair of jets, or a \PZ boson or a photon recoiling against a jet, or several jets, are used to account for any residual differences in jet energy scale in data and simulation. Additional selection criteria are applied to each jet to remove jets potentially dominated by anomalous contributions from various subdetector components. After all calibrations, the jet \pt is found from simulation to be within 5--10\% of the true \pt of the clustered particles, over the measured range~\cite{Khachatryan:2016kdb,CMS-DP-2016-020}.

For the AK8 jet mass measurement, the ``pileup per  particle identification'' algorithm~\cite{PUPPI} (PUPPI) is applied to remove pileup effects from the jet.
Particles from the PF algorithm, with their PUPPI weights, are clustered into AK8-PUPPI jets which are groomed~\cite{Salam:2009jx} to remove soft and wide-angle radiation using the soft-drop algorithm~\cite{Dasgupta:2013ihk,Larkoski:2014wba}, using the soft radiation fraction parameter $z=0.1$ and the angular exponent parameter $\beta=0$. Dedicated mass corrections~\cite{CMS-PAS-JME-16-003,CMS-B2G-16-026}, derived from simulation and data in a region enriched with \ttbar events containing merged $\PW \to \qqbar$ decays, are applied to the jet mass in order to remove residual dependence on the jet \pt, and to match the jet mass scale and resolution observed in data.
The AK8 jet soft-drop mass is assigned by matching the groomed AK8-PUPPI jet with the original jet using the criterion $\Delta R(\text{AK8 jet, AK8-PUPPI jet}) < 0.8$, where $\Delta R \equiv \sqrt{\smash[b]{(\Delta\eta)^2 + (\Delta\phi)^2}}$, $\phi$ being the azimuthal angle in radians. The matching efficiency is 100\% in the selected event sample.

\section{Event simulation\label{sec:Simulation}}

{\tolerance=800 The bulk graviton and radion signal events are simulated at leading order in the mass range 750--3000\GeV with a width of 1\MeV (much smaller than experimental resolution), using the {\MGvATNLO~2.3.3}~\cite{MG5_aMCNLO} event generator. The {NNPDF3.0} leading order parton distribution function (PDF) set~\cite{Ball:2014uwa}, taken from {LHAPDF6} PDF set~\cite{Harland-Lang:2014zoa,Buckley:2014ana,Carrazza:2015hva,Butterworth:2015oua}, with the four-flavour scheme, is used.  The showering and hadronization of partons are simulated with \PYTHIA~8.212~\cite{Pythia8p2}. \par}
The \HERWIGpp~2.7.1~\cite{Bahr:2008pv} generator is used as an
alternative model, to evaluate the systematic uncertainty
associated with the parton shower and hadronization.
The tune {\sc CUETP8M1-NNPDF2.3LO}~\cite{CUETTune} is used for \PYTHIA~8,
while the EE5C tune~\cite{HerwigppEE5C} is used for \HERWIGpp.

Non-resonant $\PH\PH$ signals were generated using the effective field theory approach defined in Refs.~\cite{deFlorian:2016spz,CarvalhoHHEFT} and is described by the five parameters given in Eq.~\ref{eq:lag}: $\kappa_{\lambda}$, $\kappa_{\cPqt}$, $c_2$, $c_{\Pg}$, and $c_{2\Pg}$.
The final state kinematic distributions of the $\PH\PH$ pairs depend upon the values of these five parameters.
A statistical approach was developed to identify twelve regions of the parameter space, referred to as clusters, with distinct kinematic observables of the $\PH\PH$ system.
In particular, models in the same cluster have similar distributions of the di-Higgs boson invariant mass \mHH, the transverse momentum of the di-Higgs boson system, and the modulus of the cosine of the polar angle of one Higgs boson with respect to the beam axis, while the distributions of these variables are unique when comparing models from different clusters~\cite{Carvalho2016}.
For each cluster, a set of representative values of the five parameters is chosen, referred to as the "shape benchmarks".
Events are simulated for each of these shape benchmarks, as well as for the SM values of these couplings, and the case where the Higgs boson self-coupling vanishes, \ie $\kappa_{\lambda}=0$.
The values of these benchmark coupling parameters are given in Table~\ref{tab:benchmarks}.

\begin{table}[htb]
\centering
\topcaption{Parameter values of the couplings corresponding to the twelve shape benchmarks, the SM prediction, and the case with vanishing Higgs boson self-interaction, $\kappa_\lambda=0$.}
\newcolumntype{.}{D{.}{.}{-2.1}}
  \begin{tabular}{r.....}
\hline
Shape benchmark & \multicolumn{1}{c}{$\kappa_{\lambda}$} & \multicolumn{1}{c}{$\kappa_{\cPqt}$} & \multicolumn{1}{c}{$c_2$} & \multicolumn{1}{c}{$c_{\Pg}$} & \multicolumn{1}{c}{$c_{2\Pg}$} \\ \hline
1 & 7.5 & 1.0 & -1.0 & 0.0 & 0.0\\
2 & 1.0 & 1.0 & 0.5 & -0.8 & 0.6 \\
3 & 1.0 & 1.0 & -1.5 & 0.0 & -0.8 \\
4 & -3.5 & 1.5 & -3.0 & 0.0 & 0.0 \\
5 & 1.0 & 1.0 & 0.0 & 0.8 & -1.0 \\
6 & 2.4 & 1.0 & 0.0 & 0.2 & -0.2 \\
7 & 5.0 & 1.0 & 0.0 & 0.2 & -0.2 \\
8 & 15.0 & 1.0 & 0.0 & -1.0 & 1.0 \\
9 & 1.0 & 1.0 & 1.0 & -0.6 & 0.6 \\
10 & 10.0 & 1.5 & -1.0 & 0.0 & 0.0 \\
11 & 2.4 & 1.0 & 0.0 & 1.0 & -1.0 \\
12 & 15.0 & 1.0 & 1.0 & 0.0 & 0.0 \\
SM & 1.0 & 1.0 & 0.0 & 0.0 & 0.0 \\
$\kappa_{\lambda}=0$ & 0.0 & 1.0 & 0.0 & 0.0 & 0.0 \\
\hline
\end{tabular}
\label{tab:benchmarks}
\end{table}

The dominant background consists of events comprised uniquely of jets (multijet events) arising from the SM quantum chromodynamics (QCD) interaction, and is modelled entirely from data. The remaining background, consisting mostly of \ttjets events, is less than 10\% of the total background, is modelled using {\POWHEG~2.0}~\cite{POWHEG, POWHEG_Frixione, POWHEGBOX} and interfaced to \PYTHIA~8. The {\sc CUETP8M2T4} tune~\cite{Skands:2014pea,CMS-PAS-TOP-17-007} is used for generating the \ttjets events.
The \ttjets background rate is estimated using a next-to-next-to-leading order cross section of $832^{+46}_{-52}\unit{pb}$~\cite{Czakon:2011xx}, corresponding to the top quark mass of 172.5\GeV.
A sample of multijet events from QCD interactions, simulated at leading order using \MGvATNLO and \PYTHIA~8, is used to develop and validate the background estimation techniques, prior to being applied to the data.

All generated samples were processed through a \GEANTfour-based~\cite{Agostinelli:2002hh,GEANT} simulation of the CMS detector. The effect of pileup, averaging to 23 at the LHC beam conditions in 2016, is included in the simulations, and the samples are reweighted to match the distribution of the number of $\Pp\Pp$ interactions observed in the data, assuming a total inelastic $\Pp\Pp$ collision cross section of 69.2\unit{mb}~\cite{CMSLumi13TeV}.

\section{Event selection\label{sec:EvtSel}}

Five different HLT triggers were used to collect the semi-resolved events used in this analysis.
An event is selected if the scalar sum of the \pt of all AK4 jets in the event (\HT) is greater than 800 or 900\GeV, depending on the LHC beam instantaneous luminosity.
Events with $\HT \ge 650\GeV$, and a pair of jets with invariant mass above 900\GeV and a pseudorapidity separation $\abs{\Delta\eta}<1.5$ are also selected.
A third HLT trigger accepts events if the scalar sum of the \pt of all AK8 jets is greater than 650 or 700\GeV and the ``trimmed mass'' of an AK8 jet is above 50\GeV. The jet trimmed mass is obtained after removing remnants of soft radiation with the jet trimming technique~\cite{Krohn:2009th}, using a subjet size parameter of 0.3 and a subjet-to-AK8 jet \pt fraction of 0.1.
Should an event contain an AK8 jet with $\pt > 360\GeV$ and a trimmed mass greater than 30\GeV, it is selected by the fourth HLT trigger.
Events containing two AK8 jets having $\pt > 280$ and 200\GeV, with at least one having trimmed mass greater than 30\GeV together with an AK4 jet passing a loose $\cPqb$-tagging criterion, pass the fifth HLT trigger.

Jets in events collected using the logical OR of the above HLT triggers are required to have $\abs{\eta}<2.4$, and $\pt > 30\GeV$ for AK4 jets and $\pt > 300\GeV$ for AK8 jets.
One AK8 jet is used to identify a boosted and spatially merged \Hbb decay (\PH jets) while two AK4 jets are used to reconstruct a spatially resolved \Hbb decay.

The first \PH-tagging criterion requires an AK8 jet to have a soft-drop mass $m_{\text{J}}$ between 105 and 135\GeV, consistent with the measured mass of the Higgs boson $\mH=125\GeV$.
This selection corresponds to an efficiency of about 60--70\% for a resonant signal mass \mx in the range 750--3000\GeV.
The soft-drop jet mass interval was chosen to include a large fraction of the boosted \Hbb signal, while avoiding overlaps with CMS analyses searching for bulk gravitons and radions decaying to boosted \PW~and \PZ bosons~\cite{CMS-B2G-16-007}.
The ``$N$-subjettiness'' algorithm~\cite{Thaler:2011gf} is used on the AK8-PUPPI jet constituents, to compute the variables $\tau_{\text{N}}$, which quantify the degree to which a jet contains $N$ subjets.
A selection on the ratio $\nsub \equiv \tau_2/\tau_1 < 0.55$ is required for all AK8 jets to be \PH tagged, which has a jet \pt-dependent efficiency of 50--70\%.
The selection criterion on $\nsub$ was optimized for signal sensitivity over the range of \mx values explored.

A jet flavour requirement using a ``\Hbbt'' algorithm~\cite{Sirunyan:2017ezt} is applied to the AK8 jet as the final \PH-tagging requirement. The \Hbbt is a multivariate discriminator with an output between -1 and 1, a higher value indicating a greater probability for the jet to contain a \bbbar pair. The \Hbbt exploits the presence of two hadronized $\cPqb$ quarks inside the boosted \Hbb decay, and uses variables related to $\cPqb$ hadron lifetime and mass to distinguish $\PH$ jets against a background of jets of other flavours. The \Hbbt algorithm also exploits the fact that the directions of the $\cPqb$ hadron momentum are strongly correlated with the axes used to calculate the $N$-subjettiness variables $\tau_{\text{N}}$.
An \PH jet candidate should have a \Hbbt discriminator greater than 0.8, which corresponds to an efficiency of 30\% and a misidentification rate of about 1\%, as measured in a sample of multijet events.
The efficiency of the \Hbbt for simulated jets is corrected to match that in the data, based on efficiency measurements using jets containing pairs of muons, thereby yielding samples enriched in jets from gluons splitting to \bbbar pairs. These efficiency corrections are in between 0.92 and 1.02, for jets in the selected \pt range.

To find a Higgs boson decay into two resolved $\cPqb$ quark jets, all AK4 jets in each event are examined for their $\cPqb$ tag value using ``DeepCSV'' algorithm, which is a deep neural network, trained using information from tracks and secondary vertices associated to the jets~\cite{Sirunyan:2017ezt}.
The DeepCSV discriminator gives the probability of a jet to have originated from the hadronization of a bottom quark. A selection on the DeepCSV discriminator of AK4 jets is made, corresponding to a 1\% mistag rate for light flavoured jets. The corresponding $\cPqb$-tagging efficiency is about 70\% for $\cPqb$ quark jets in the \pt range 80--150\GeV, and decreases to about 50\% for $\pt\!\sim\!1000\GeV$. The $\cPqb$ tagging efficiency in the simulations is corrected to match the one in the data, using measurements of the $\cPqb$ tagging algorithm performance in a sample of muon-tagged jets and $\cPqb$ jets from \ttjets events, where the correction factor ranges from approximately 0.95 to 1.1.

To identify events with a resolved \Hbb decay, all pairs of $\cPqb$-tagged AK4 jet are examined, to find events with at least one pair where each AK4 jet is at least $\Delta R >$ 0.8 away from the leading-\pt AK8 jet and within $\Delta R < 1.5$ of each other. If several such pairs are found, the pair of jets, $\text{j}_{1}$ and $\text{j}_{2}$, that has the highest sum of the AK4 jet DeepCSV discriminator values is selected. The leading-\pt AK8 jet is then identified as the boosted \PH candidate, and the pair of AK4 jets is identified as the resolved \PH candidate. If no pairs are found, this process is repeated with the subleading-\pt AK8 jet. If a pair of AK4 jets is identified, then the subleading-\pt AK8 jet is identified as the boosted \PH candidate, and the pair of AK4 jets is identified as the resolved \PH candidate. If no pairs are found once again, the event is rejected.
The invariant mass of $\text{j}_{1}$ and $\text{j}_{2}$, $\mjj(\text{j}_{1}, \text{j}_{2})$, is required to be within 90--140\GeV, forming the resolved \Hbb candidate.

The \ttjets background is reduced by reconstructing a $\cPqt\to \cPqb\cPq\cPaq$ system in events with three or more AK4 jets, combining $\text{j}_{1}$ and $\text{j}_{2}$ with the nearest AK4 jet $\text{j}_{3}$.
For the \ttjets background, the trijet invariant mass $m_{\text{jjj}}(\text{j}_{1}, \text{j}_{2}, \text{j}_{3})$ peaks around the top quark mass of 172\GeV.
Hence, $m_{\text{jjj}}(\text{j}_{1}, \text{j}_{2}, \text{j}_{3})$ is required be greater than 200\GeV, namely above the top quark mass.
Events containing leptons (electrons or muons) with $\pt > 20\GeV$ and $\abs{\eta}<2.4$, and only a small amount of energy in an area around the lepton direction compared to the lepton \pt, are rejected, to further suppress \ttjets and other backgrounds.

A resonant $\PH\PH$ signal results in a small pseudorapidity separation between the two Higgs bosons, while the candidates from the multijet background typically have a larger pseudorapidity separation.
Events are therefore categorized according to the pseudorapidity difference between the \PH jet and the resolved \Hbb candidate. These two categories are defined by $\abs{\Delta\eta(\text{\PH jet, resolved \Hbb})}$ within the interval 0.0--1.0 or within the interval 1.0--2.0.

To search for resonant and a non-resonant $\PH\PH$ signals, the invariant mass distribution of the boosted and resolved Higgs boson candidate system ($m_{\text{Jjj}}$) is examined for an excess of events over the estimated background.
The ``reduced di-Higgs invariant mass'' is defined as $\mjjjs \equiv m_{\text{Jjj}} - (m_{\text{J}} - \mH) - (\mjj(\text{j}_{1}, \text{j}_{2}) - \mH)$. The quantity \mjjjs is used rather than $m_{\text{Jjj}}$ since by subtracting the masses of the reconstructed \PH candidates and adding back the exact Higgs boson mass \mH, fluctuations from the jet mass resolution are corrected, leading to 8--10\% improvement in the $\PH\PH$ mass resolution. After the full selection, the multijet background is about 90\% of the total background, with the remaining background being \ttjets events.

With the above event selection, the trigger criteria reach an efficiency of greater than 99\% for events with $\mjjjs \ge 1100\GeV$. For lower values of invariant mass (between 750 and 1100\GeV), the trigger efficiency is between 80 and 99\% for $0 \le \abs{\Delta\eta(\text{$\PH$ jet, resolved \Hbb})} < 1$ and between 60 and 99\% for $1 \le \abs{\Delta\eta(\text{$\PH$ jet, resolved \Hbb})} \le 2$.
 The trigger efficiency for the data is estimated from a multijets sample collected with a control trigger requiring a single AK4 jet with $\pt > 260\GeV$. The trigger efficiency for the simulated samples is corrected using a scale factor to match the observed efficiency for the data. This scale factor depends mildly on $\abs{\Delta\eta(\text{$\PH$ jet, resolved \Hbb})}$, and is hence applied as a function of this variable.

\begin{figure}[h]
\centering
\includegraphics[width=0.49\textwidth]{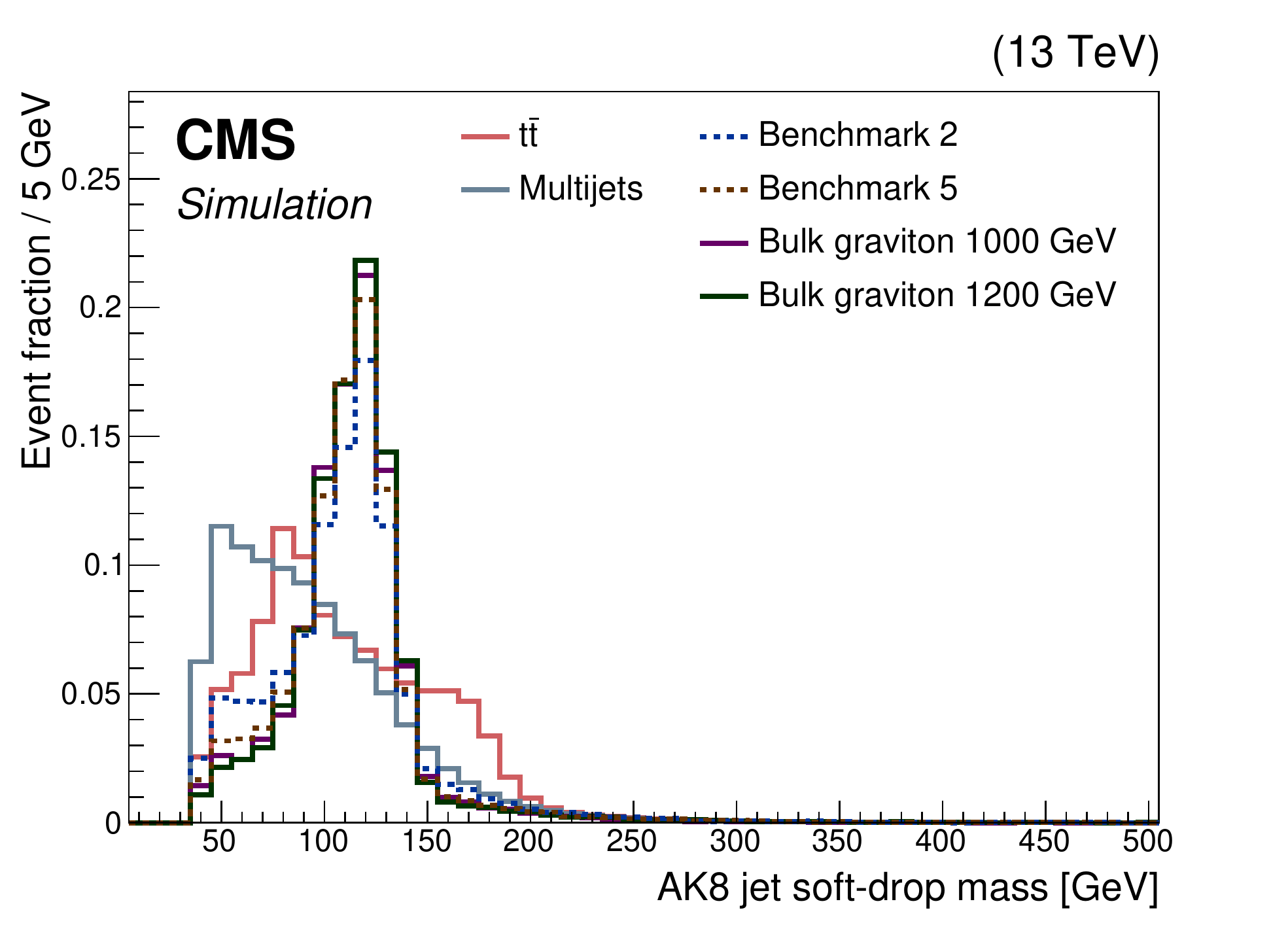}
\includegraphics[width=0.49\textwidth]{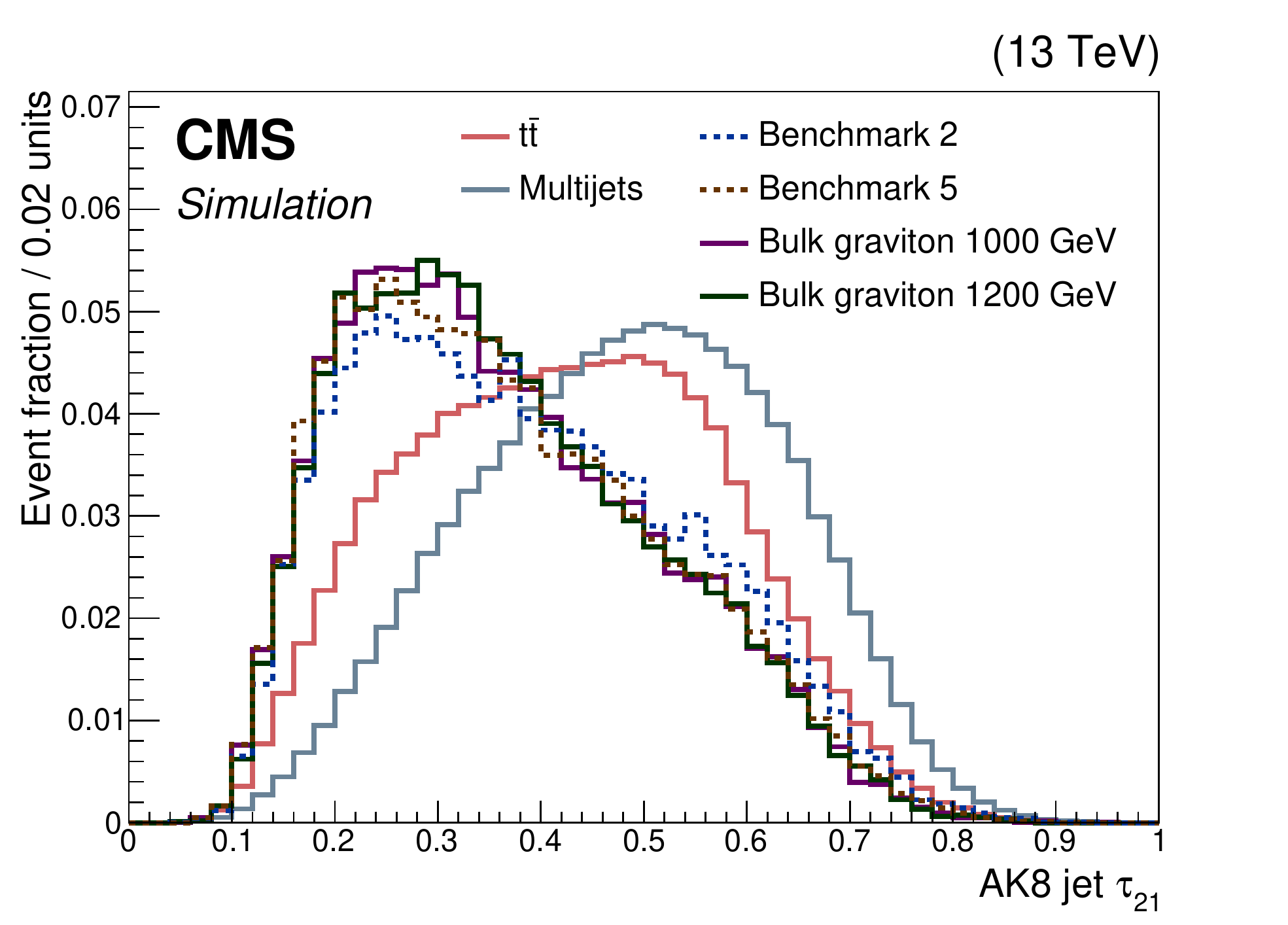}
\includegraphics[width=0.49\textwidth]{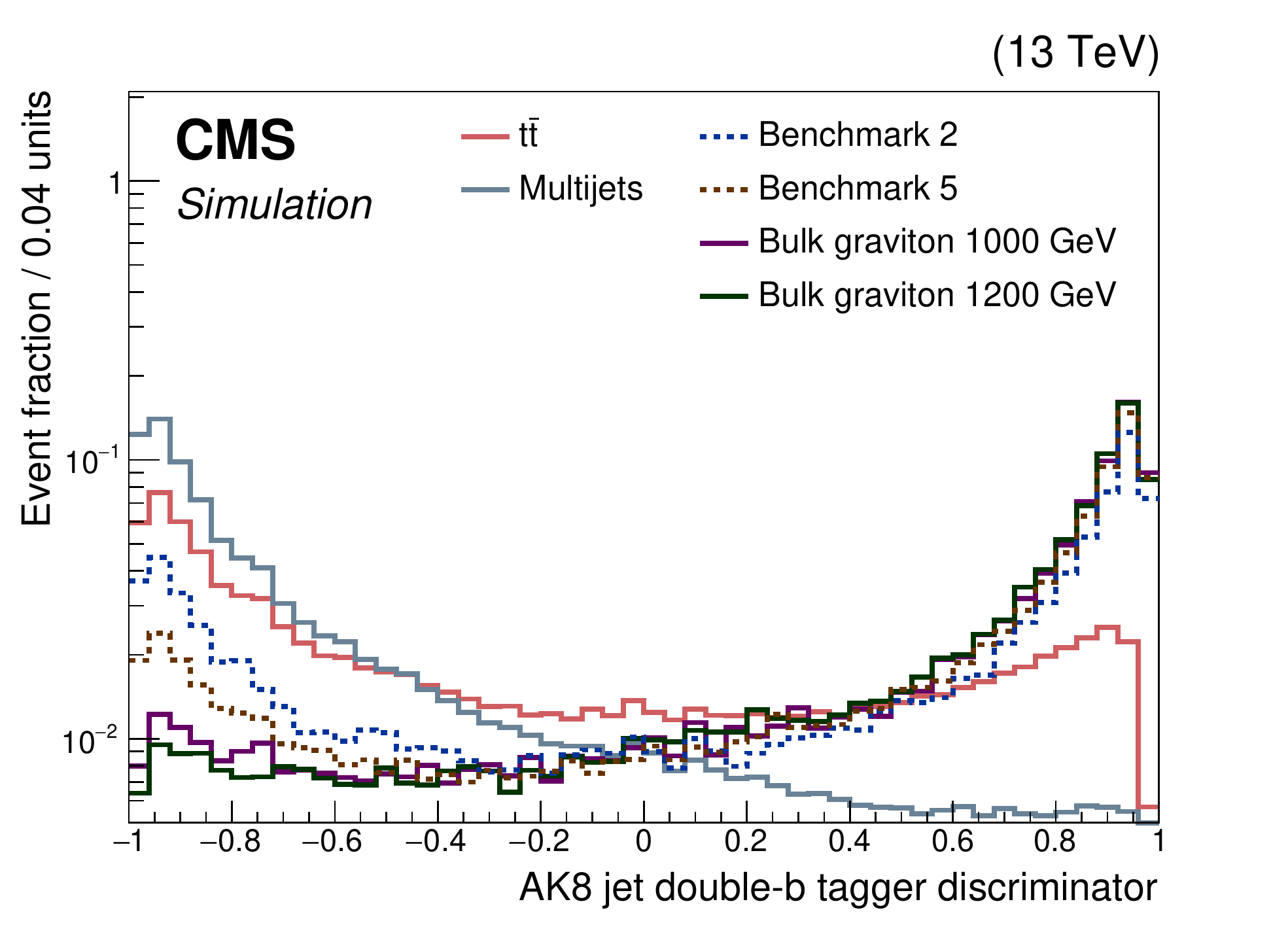}
\caption{Distributions of the soft-drop mass (upper left), \nsub (upper right), and the \Hbbt (lower), for AK8 jets in semi-resolved events. The multijet and the \ttjets background components are shown separately, along with the simulated signals for bulk gravitons of masses 1000 and 1200\GeV and the non-resonant benchmark models 2 and 5. The distributions are normalized to unity.\label{fig:AK8substr_presel}}
\end{figure}

\begin{figure}[h]
\centering
\includegraphics[width=0.49\textwidth]{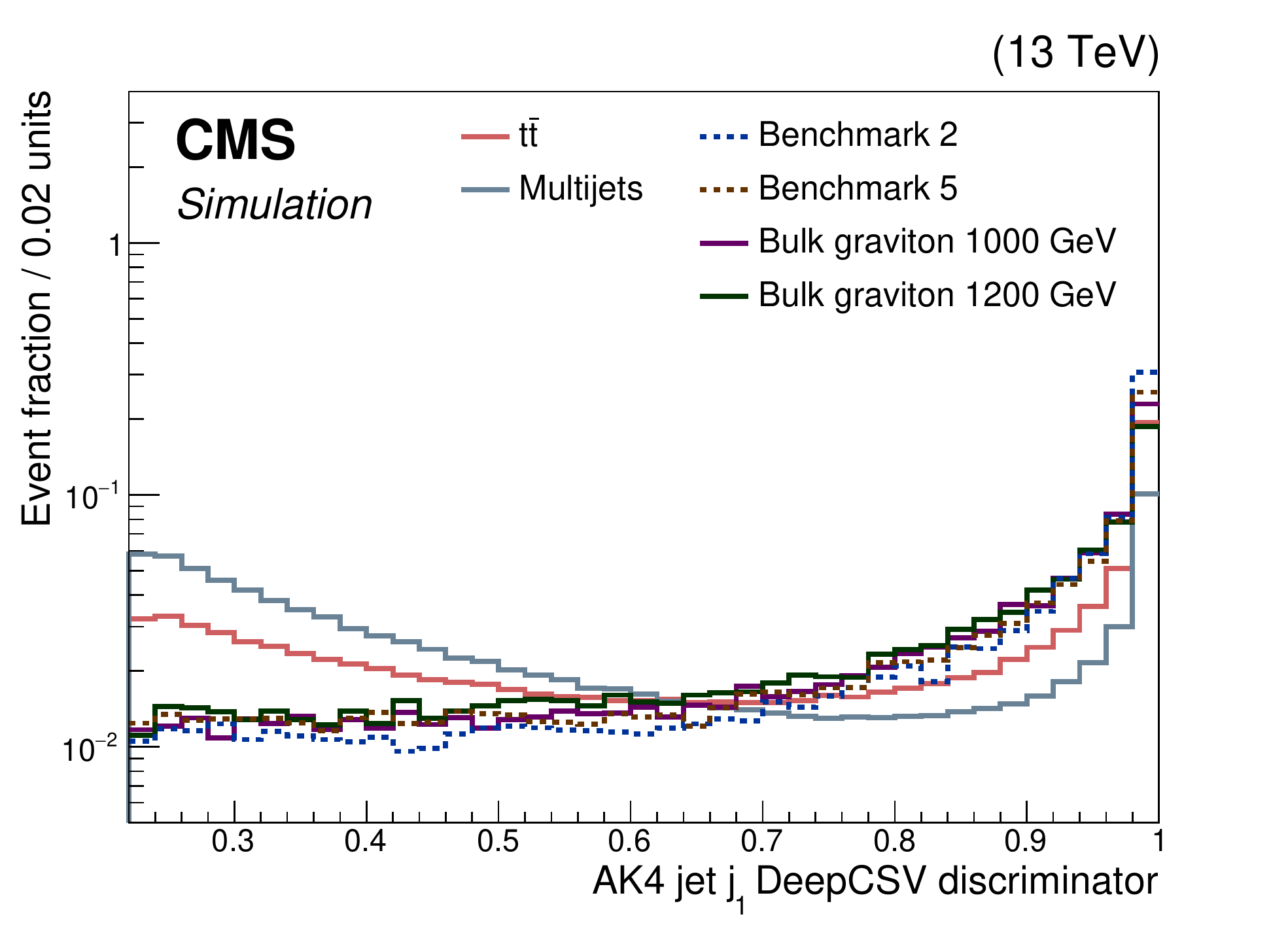}
\includegraphics[width=0.49\textwidth]{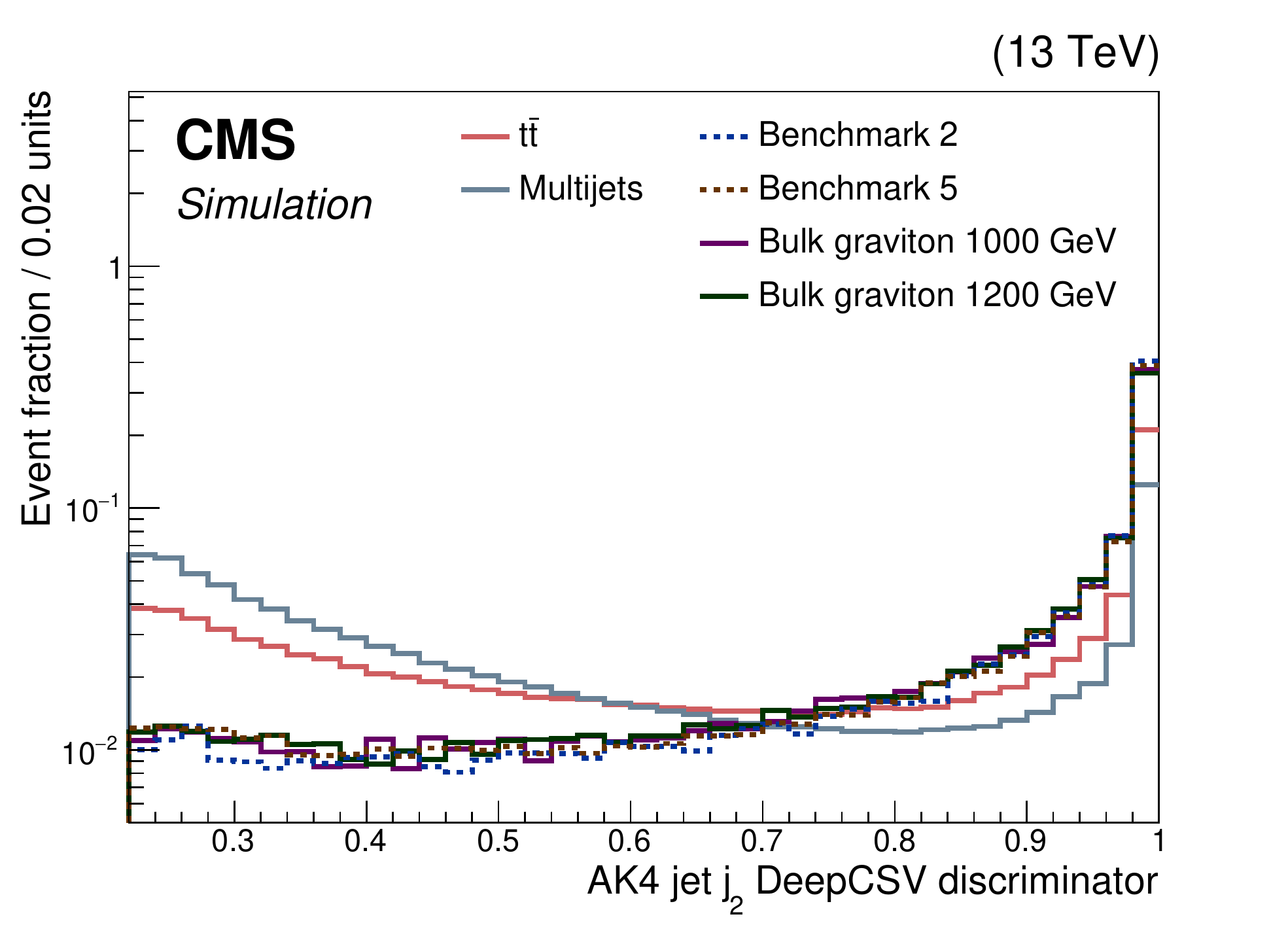}
\includegraphics[width=0.49\textwidth]{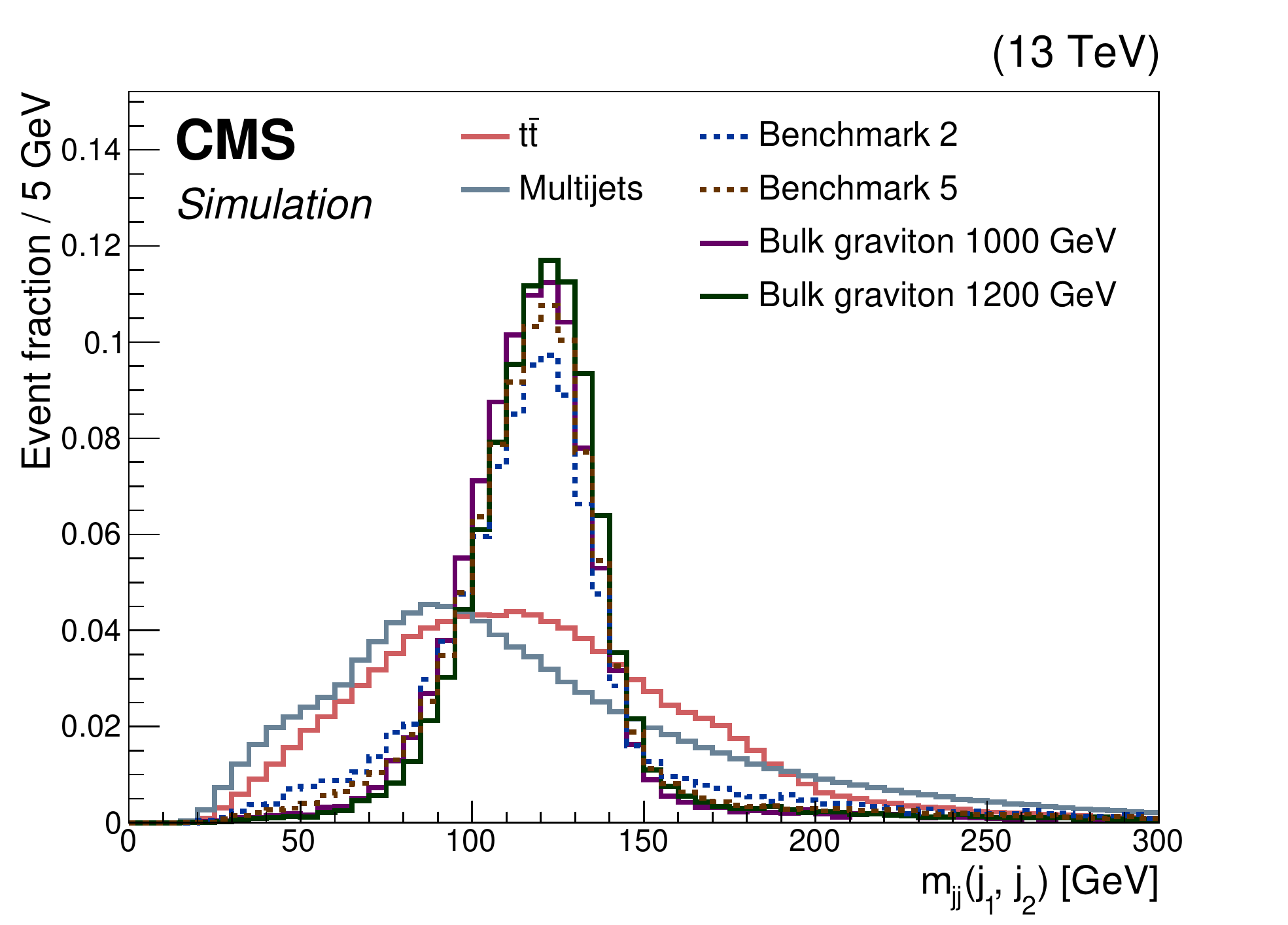}
\includegraphics[width=0.49\textwidth]{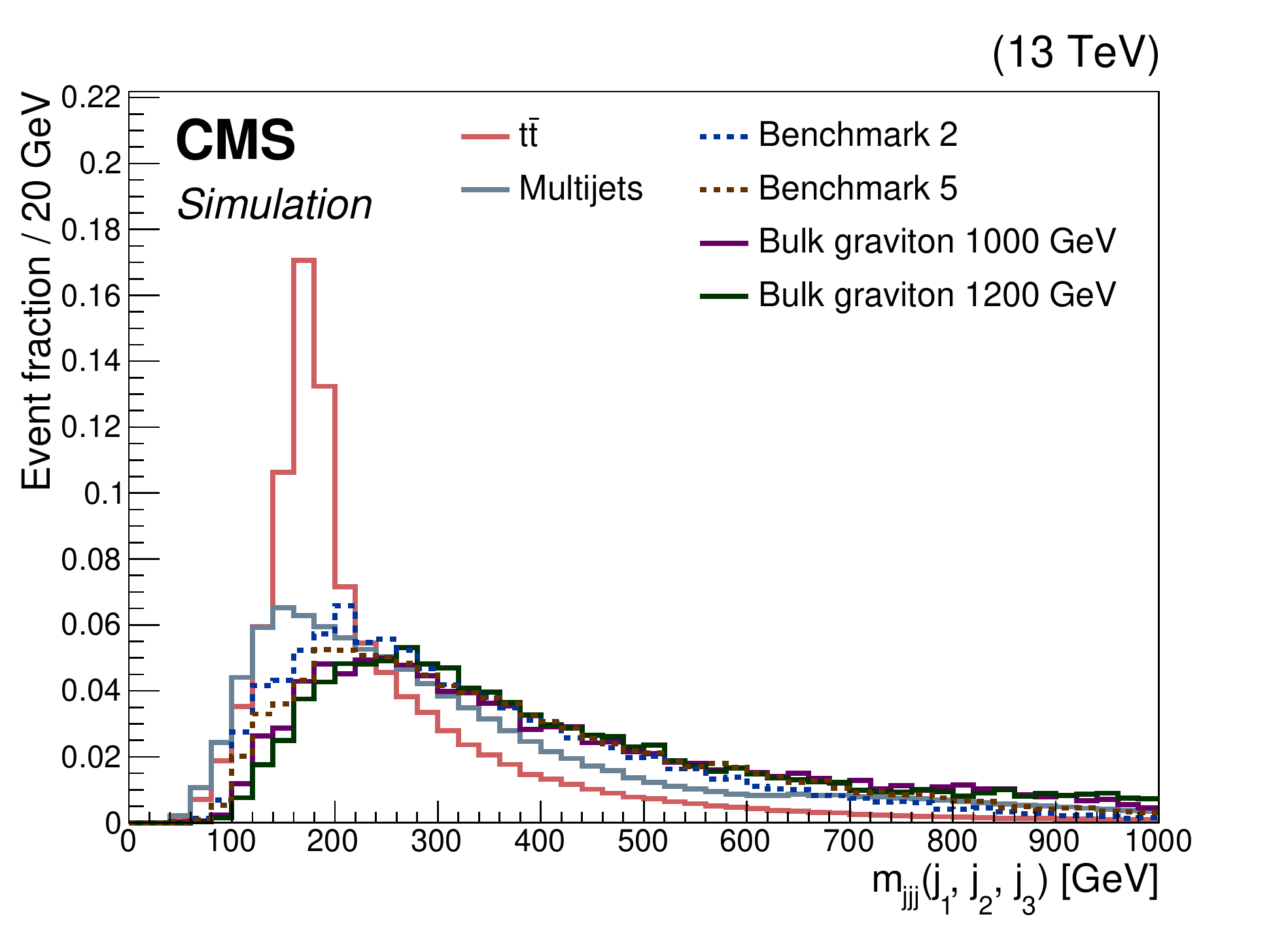}
\caption{Distributions for AK4 jets of the DeepCSV discriminators for the leading $\text{j}_{1}$ (upper left) and next leading $\text{j}_{2}$ (upper right), the invariant mass of $\text{j}_{1}$ and $\text{j}_{2}$, $\mjj(\text{j}_{1}, \text{j}_{2})$ (lower left), and the invariant mass of $\text{j}_{1}$, $\text{j}_{2}$, and their nearest AK4 jet $\text{j}_{3}$, $\mjjj(\text{j}_{1}, \text{j}_{2}, \text{j}_{3})$ (lower right), in semi-resolved events. The multijet and \ttjets background components are shown separately, along with the simulated signals for bulk gravitons of masses 1000 and 1200\GeV and the non-resonant benchmark models 2 and 5. The distributions are normalized to unity.}
\label{fig:AK4vars_presel}
\end{figure}

The AK8 jet soft-drop mass distribution, the $N$-subjettiness ratio \nsub distribution, and the \Hbbt discriminator distribution for the backgrounds and simulated signals are shown in Fig.~\ref{fig:AK8substr_presel}.
The DeepCSV discriminator distributions for the two AK4 jets, the dijet invariant mass distribution, and the trijet invariant mass distribution for the backgrounds and simulated samples are shown in Fig.~\ref{fig:AK4vars_presel}. The selection criteria for the above plots is as follows:
AK8 jets with $\pt > 300\GeV$, AK4 jets with $\pt > 30\GeV$, AK8 and AK4 jets with $\abs{\eta}<2.4$, AK8 jet soft-drop mass $> 40\GeV$, AK4 jets DeepCSV discriminator $> 0.2219$, $\Delta R < 1.5$ separation between the AK4 jets, and $\Delta R > 0.8$ separation between the AK8 jet and each AK4 jet.

\begin{table}[h]
  \centering
  \topcaption{Summary of the offline selection criteria for semi-resolved $\PH\PH\to\bbbar\bbbar$ events.}
  \label{tab:evsel}
    \begin{tabular}{lc}
      \hline
      Variable & Selection  \\
      \hline
      At least 1 AK8 jet J  & $\pt > 300\GeV$, $\abs{\eta}<2.4$\\
      At least 2 AK4 jets $\text{j}_{1}$ and $\text{j}_{2}$ & $\pt > 30\GeV$, $\abs{\eta}<2.4$\\
      $\Delta R(\text{J},\text{j}_{\text{i}})$ & ${>}0.8$\\
      $\Delta R$($\text{j}_{1}$,$\text{j}_{2}$) & ${<}1.5$\\
      $\abs{\Delta\eta$(J,$\text{j}_{1}$+$\text{j}_{2}$)$}$ & ${\le}2$\\
      \mjjjs & ${>}750\GeV$\\
      J soft-drop mass & 105--135\GeV\\
      J $\tau_{21}$ & ${<}0.55$\\
      J \Hbbt discriminator & ${>}0.8$ \\
      $\text{j}_{1}$+$\text{j}_{2}$ mass & 90--140\GeV\\
      $\text{j}_{1}$+$\text{j}_{2}$+(nearest AK4 jet) mass & ${>}200\GeV$\\
      $\text{j}_{1}$ and $\text{j}_{2}$ DeepCSV & 70\% $\cPqb$-tagging eff., 1\% mistag \\
      Number of isolated leptons ($\Pe$ or $\mu$) & ${=}0$ \\
      \hline
    \end{tabular}
\end{table}

\begin{figure}[h]
\centering
\includegraphics[width=0.49\textwidth]{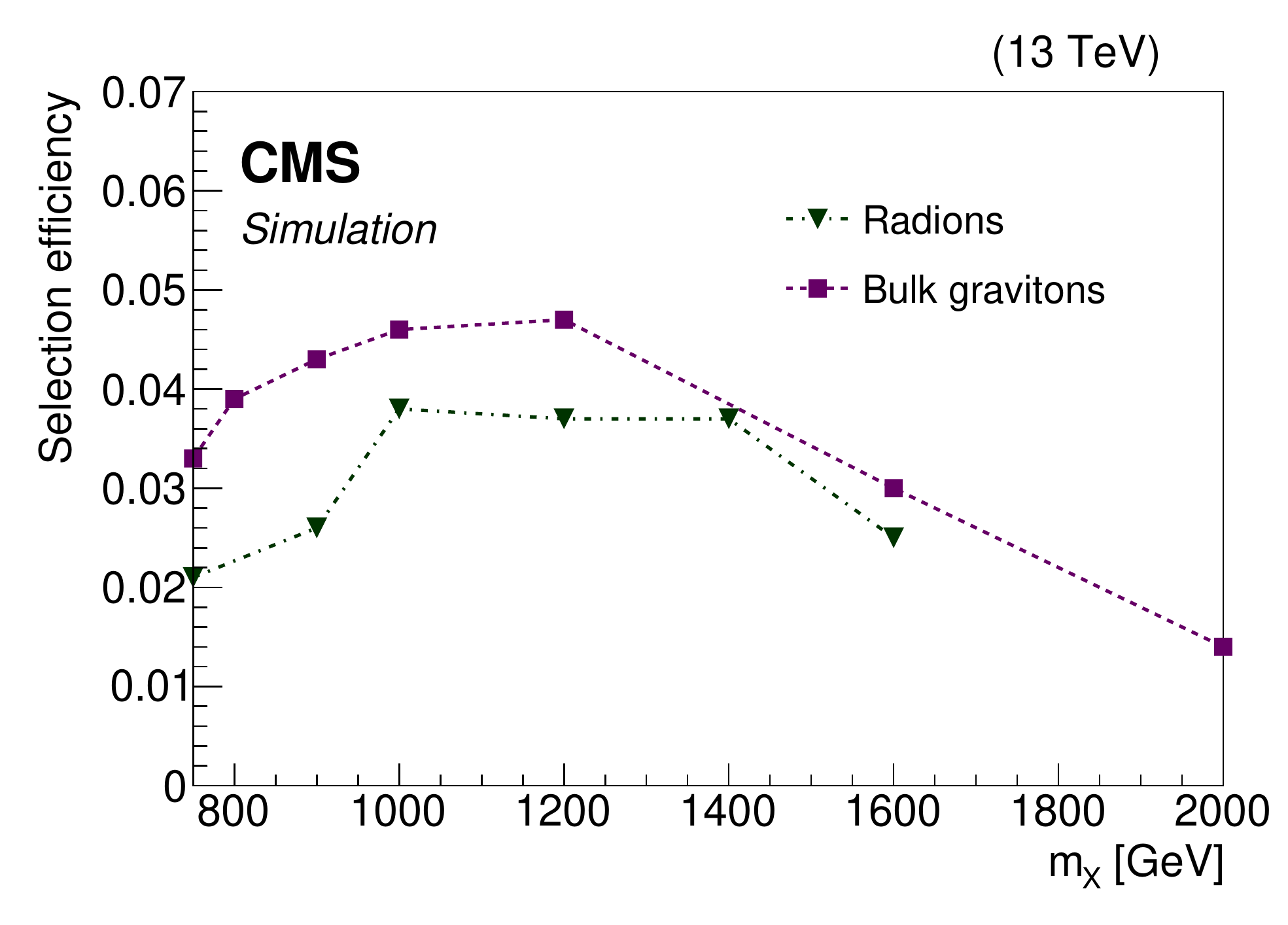}
\caption{The signal selection efficiencies for the radion and the bulk graviton, for different masses. The events are required to pass the selections given in Table~\ref{tab:evsel} as well as to fail the selections of the fully-merged analysis of Ref.~\cite{CMS-B2G-16-026}.}
\label{fig:acceptance}
\end{figure}

The semi-resolved event selection is summarized in Table~\ref{tab:evsel}, where in addition to these criteria, the events that are used by the fully-merged analysis of Ref.~\cite{CMS-B2G-16-026} are removed, as detailed at the end of this section.
The event selection efficiencies for bulk gravitons and radions are given in Fig.~\ref{fig:acceptance}, for different assumed masses in the range 750--2000\GeV.
At low masses, the efficiency rise is mainly due to the increases in the trigger efficiency and in the efficiency of the requirement on the \abs{\Delta\eta} between the two Higgs boson candidates. The latter efficiency is more important for the radion, which being a spin-0 particle has a wider \abs{\Delta\eta} at low masses than the spin-2 bulk graviton.  At high masses, the efficiency drops because more events migrate to the fully-merged regime.
The selection efficiencies for the non-resonant signals are between 0.01--2\%.

In view of the statistical combination of the semi-resolved and the fully-merged analyses, we briefly describe the search in the fully-merged topology~\cite{CMS-B2G-16-026}. The analysis in the fully-merged regime uses the same trigger selection and the same selection for the \PH jet identification, except for a different requirement on the \Hbbt. These events contain two \PH jets $\text{J}_{1}$ and $\text{J}_{2}$ instead of one. The fully-merged events are classified according to the values of the \Hbbt discriminators of the two \PH jets, with both $\text{J}_{1}$ and $\text{J}_{2}$ required to pass a loose \Hbbt discriminator value of $> 0.3$. Events are then categorized into those with both $\text{J}_{1}$ and $\text{J}_{2}$ passing a tighter \Hbbt discriminator requirement of $> 0.8$, and the rest.
The pseudorapidity separation between $\text{J}_{1}$ and $\text{J}_{2}$ is required to be $\abs{\Delta\eta(\text{J}_{1}, \text{J}_{2})} < 1.3$.
The reduced di-Higgs invariant mass for fully-merged events is defined as $m_{\text{JJ,red}}=m_{\text{JJ}} - (m_{\text{J}_{1}} -\mH) - (m_{\text{J}_{2}} - \mH)$,  where $m_{\text{JJ}}$ is the invariant mass of $\text{J}_{1}$ and $\text{J}_{2}$ and $m_{\text{J}_{1}}$ and $m_{\text{J}_{2}}$ are their soft-drop masses, respectively.

A Higgs boson candidate which passes the boosted AK8 jet selection can also pass the selection for two resolved AK4 jets. In particular, signal samples with higher mass that pass the semi-resolved selection often pass the fully-merged selection because both Higgs candidates are merged, but one candidate still passes the selection for a resolved jet as well. For each signal, the final semi-resolved selection includes anywhere from 23--53\% events that are used by the fully-merged analysis, whether in the signal region or to estimate the QCD multijets background. These events are then removed from the semi-resolved analysis to allow for a combination with the fully-merged analysis.

\section{Background estimation\label{sec:SigModelBkgEst}}

\begin{figure}[!htb]
\centering
\includegraphics[width=0.45\textwidth]{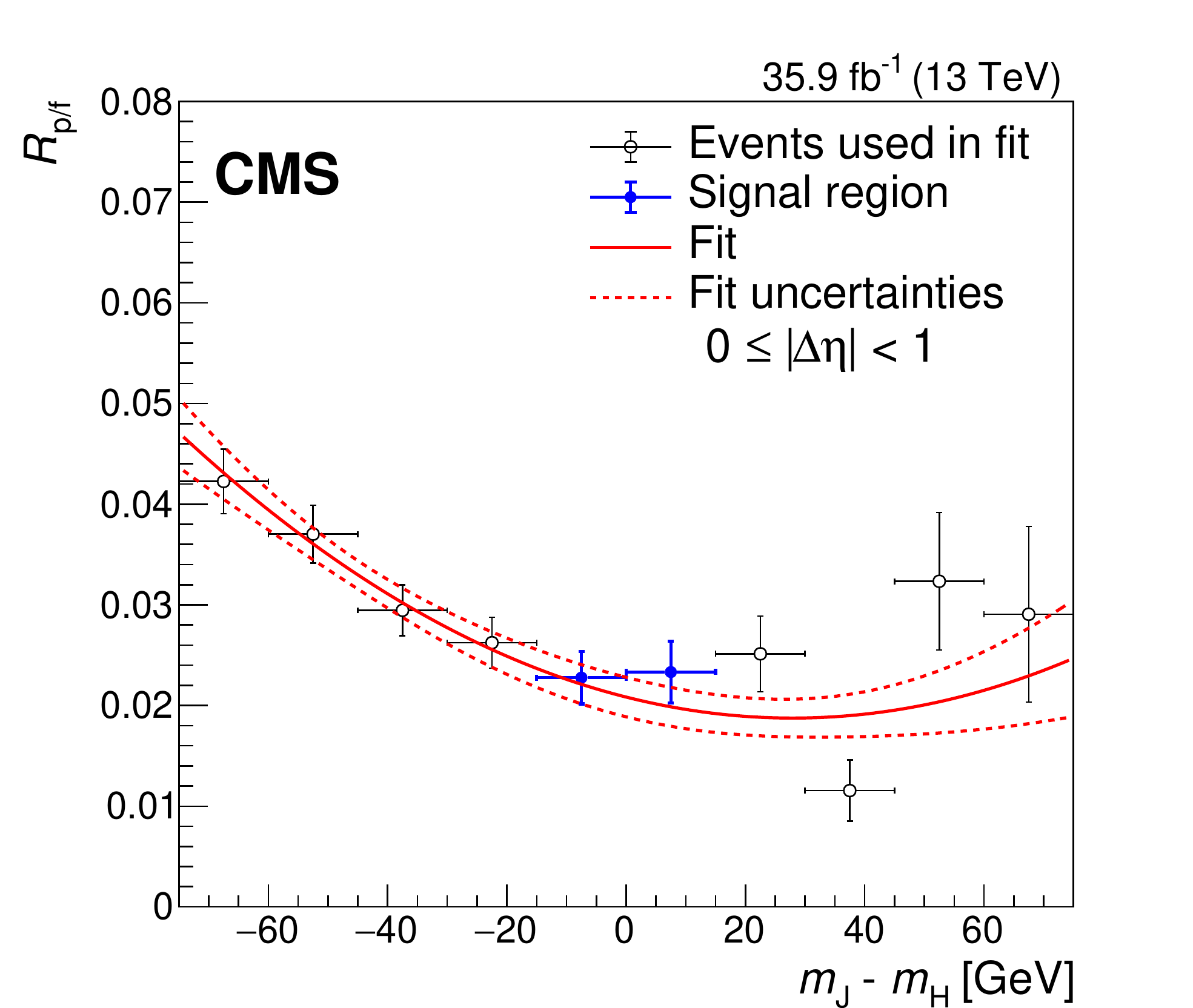}
\includegraphics[width=0.45\textwidth]{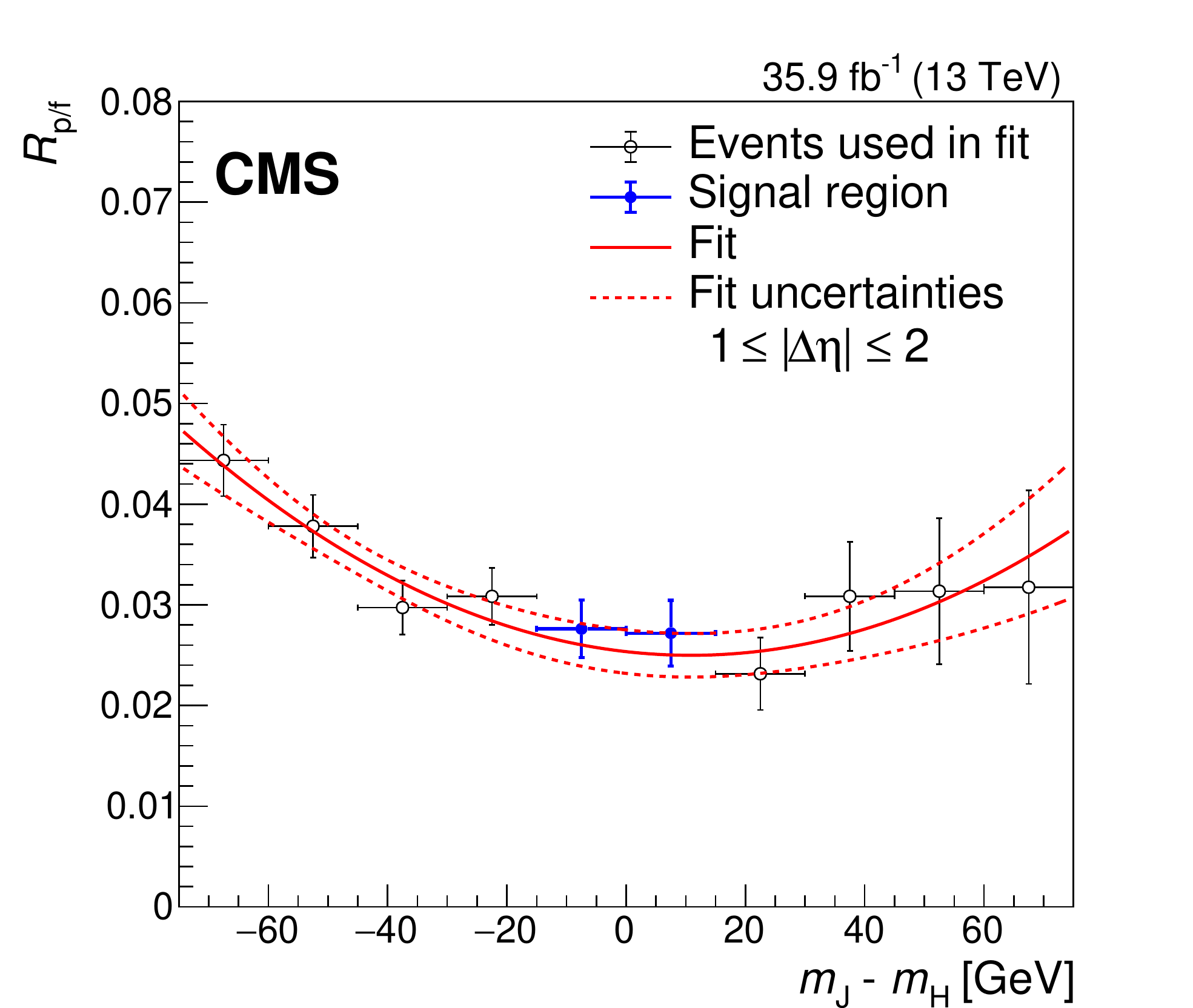}\\
\includegraphics[width=0.45\textwidth]{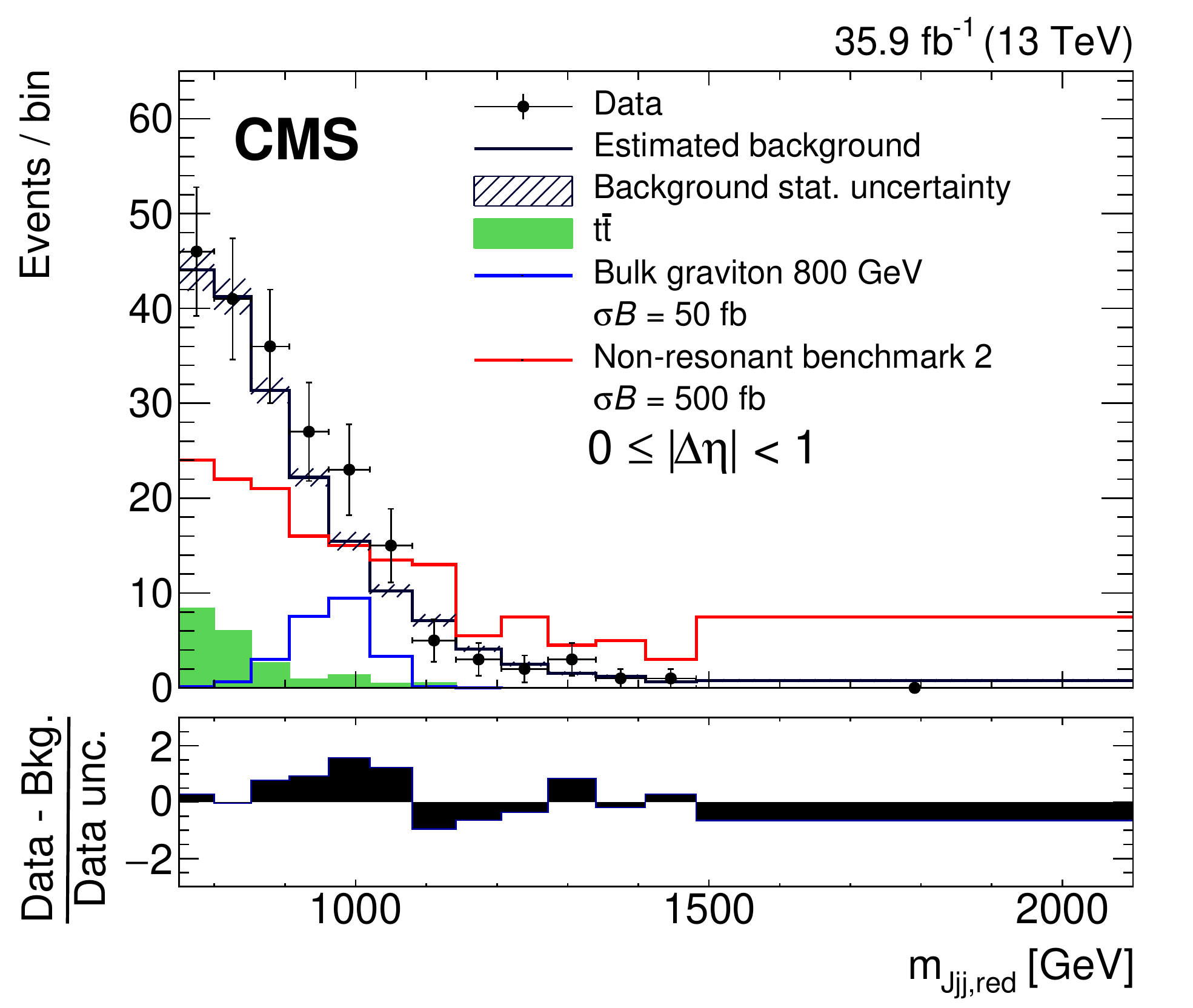}
\includegraphics[width=0.45\textwidth]{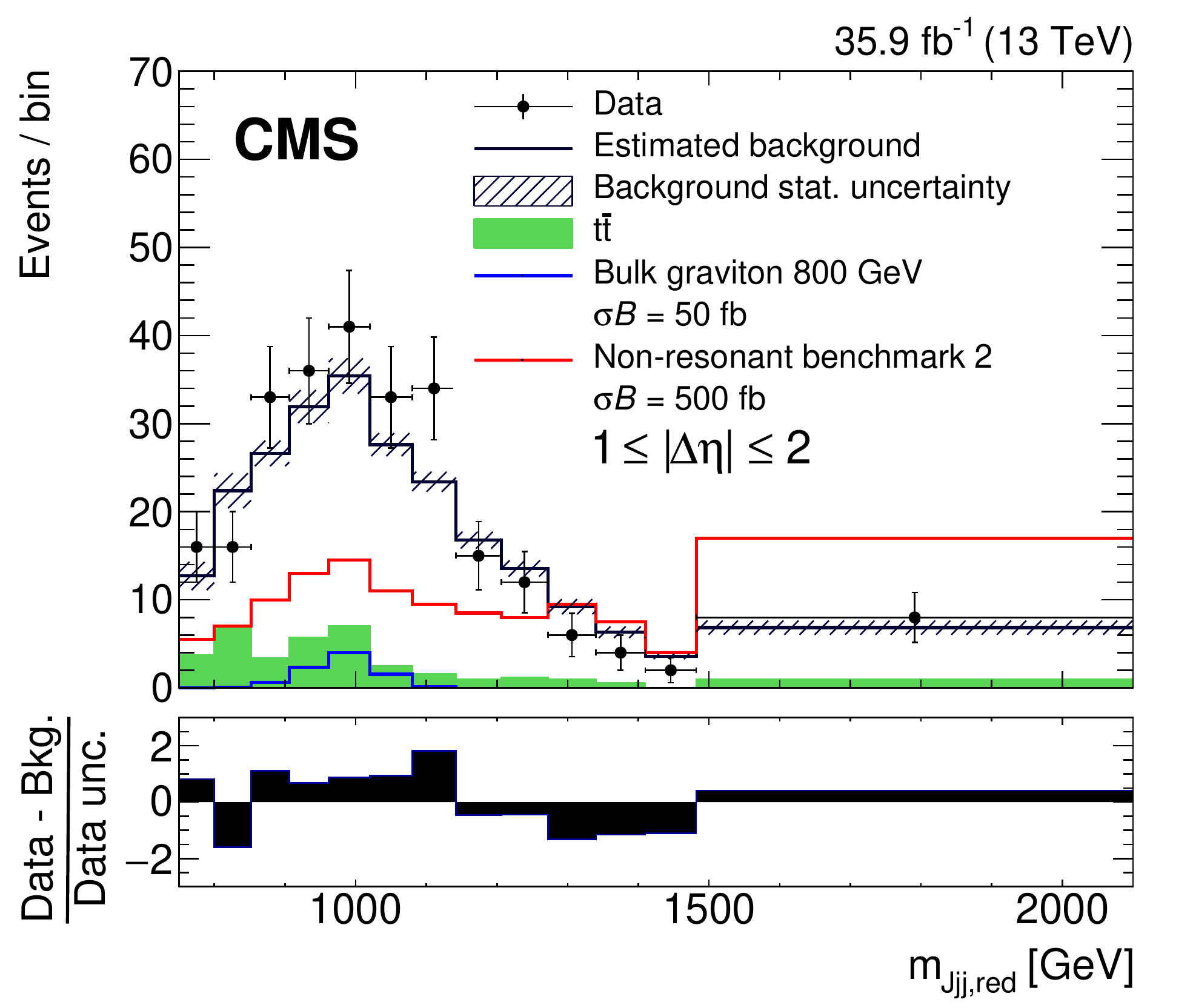}\\
  \caption{Upper: The \Hbbt pass-fail ratio $R_\text{p/f}$ of the leading-\pt AK8 jet in semi-resolved events as a function of the difference between the soft-drop mass and the Higgs boson mass, $m_{\text{J}} - \mH$. The measured ratio in different bins of $m_{\text{J}} - \mH$ is used in the fit (red solid line), except in the region around $m_{\text{J}}-\mH=0$, which corresponds to the signal region (blue markers). The fitted function is interpolated to obtain $R_\text{p/f}$ in the signal region. Lower: The reduced mass distribution \mjjjs in the data (black markers) with the estimated background represented as the black histogram. The \ttjets contribution from simulation is represented in green. The rest of the background is multijets, calculated by applying the $R_\text{p/f}$ to the antitag region. The uncertainty in the total background, before fitting the background model to the data, is depicted using the shaded region. The signal distributions for a  bulk graviton with a mass of 800\GeV (blue) and the non-resonant benchmark 2 model (red) are also shown for assumed values of the products of the production cross sections for $\PH\PH$ and the branching fraction to $4\cPqb$, $\sigma \mathcal{B}$. For the left and right figures, the pseudorapidity intervals are $0\le\abs{\Delta\eta}<1$ and $1\le\abs{\Delta\eta}\le2$, respectively.}
\label{fig:Alphabet_Rpf_semiresolved_rejectBoosted}
\end{figure}

The multijet background estimation technique for the semi-resolved analysis is the same as that for the fully-merged analysis~\cite{CMS-B2G-16-026}. A set of signal-free control regions is defined by changing the criteria on the  soft-drop mass and the \Hbbt discriminator of the selected AK8 jet from those used for the \PH tagging. The selection criteria applied to the AK4 jets forming the resolved \Hbb are the same as those used for the signal regions. If the soft-drop mass is within 60\GeV above or below the \PH jet mass window of 105--135\GeV, these regions are referred to as the mass sideband regions. These sidebands are separated into regions that pass or fail the \Hbbt tagging requirement.

We define the pass-fail ratio $R_\text{p/f}$ as the ratio of events for which the AK8 jet passes and fails the \Hbbt tagging requirement. The $R_\text{p/f}$ is measured in the soft-drop mass sidebands as a function of soft-drop mass. These values are fit to a quadratic function of the \PH jet mass to calculate the $R_\text{p/f}$ in the signal region. The antitag region, defined with the same criteria as the signal region, but with the AK8 jet failing the \Hbbt requirement, is then scaled by the $R_\text{p/f}$ value to estimate the multijets background in the signal region. This is done in bins of soft-drop mass to predict the entire background shape. The dependence of $R_\text{p/f}$ on \mjjjs was found to be negligible, within the measurement uncertainties.
Both the shape of the background \mjjjs distribution and its total yield in the signal region is obtained using this method.

Prior to estimating the background, the \ttjets contributions derived from Monte Carlo simulation are subtracted from all sideband and signal regions in the data, and then added back in once the multijet background calculation is completed, to estimate the contribution of \ttjets to the total background.
The fractions of signal events in the sideband regions were found to be negligible as compared with the total numbers of events.

Figure~\ref{fig:Alphabet_Rpf_semiresolved_rejectBoosted} (left) shows the quadratic fit in the
AK8 jet soft-drop mass sidebands of the pass-fail ratio $R_\text{p/f}$ as a function of
AK8 jet soft-drop mass, as obtained in the data and in the predicted background shape in the signal region, where overlap with the merged analysis in the signal, sideband, and antitag regions is removed. A $\chi^{2}$ test statistic was used to perform the fit, and the modelling was validated using Monte Carlo simulations and control samples in the data.
The functional form was chosen after performing a Fisher F-test~\cite{Fisher}, which established that, among polynomials, a quadratic form is necessary and sufficient. Other functional forms were tested and the fit results were found to be consistent with that using the quadratic function.
The resulting background distributions are compared with the observed data, as shown in Fig.~\ref{fig:Alphabet_Rpf_semiresolved_rejectBoosted} (right).

\section{Systematic uncertainties\label{sec:Systematics}}

The following sources of systematic uncertainty affect the expected
signal and background event yields. None of these lead to a
significant change in the signal shape. A complete list of systematic uncertainties is given in Table~\ref{tab:Syst}.

\begin{table}[h]
  \centering
  \topcaption{Summary of the ranges of systematic uncertainties in the signal and background yields, for both the semi-resolved analysis and for the fully-merged analysis, taken from Ref.~\cite{CMS-B2G-16-026}.}
  \label{tab:Syst}
    {\renewcommand{\arraystretch}{1.2}%
    \begin{tabular}{lcc}
      \hline
      Source & Uncertainty (Semi-resolved) & Uncertainty (Fully-merged) \\
      \hline
      \multicolumn{3}{c}{Signal yield (\%)} \\
      Trigger efficiency                & 1--15    & 1--15    \\
      Jet energy scale and resolution   & 1--3     & 1        \\
      Jet mass scale and resolution     & 2        & 2        \\
      \PH tagging correction factor     & 5--20    & 7--20    \\
      \PH jet \nsub selection           & +14/-13  & +30/-26  \\
      $\cPqb$ tagging selection         & 2--9     & 2--5     \\
      PDF and scales                    & 0.1--3   & 0.1--2   \\
      Pileup modelling                  & 1--2     & 2        \\
      Luminosity                        & 2.5      & 2.5      \\
      Trijet Invariant Mass             & 0.5      & \NA      \\
      \multicolumn{3}{c}{Background yield (\%)} \\
      \ttjets cross section             & 5        & \NA      \\
      QCD background $R_\text{p/f}$ fit & 2--10    & 2--7     \\
     \hline
    \end{tabular}
    }
\end{table}

The trigger response modelling uncertainties are
particularly important for $\mjjjs < 1100\GeV$, where the trigger
efficiency drops below 99\%.
The trigger efficiency data-to-simulations scale factor has an uncertainty between 1 and 15\%, attributable to the control trigger inefficiency and the sample size used.

The impact of the jet energy scale and resolution uncertainties~\cite{CMS-PAS-JME-16-003} on the signal yields was estimated to be 1--3\%, depending on the signal mass.
The jet mass scale and resolution, as well as the $\tau_{21}$ selection efficiency data-to-simulation scale factors were measured using a sample of boosted $\PW \to {\qqbar'}$ jets in semileptonic \ttbar events.
The jet mass scale and resolution has a 2\% effect on the signal yields because of a change in the mean of the \PH jet mass distribution.
A correction factor is applied to account for the difference in the jet shower profile of $\PW \to {\qqbar'}$ and \Hbb decays, by comparing the ratio of the efficiency of \PH and \PW jets using the \PYTHIA~8 and \HERWIG{++} shower generators.
This uncertainty, the \PH tagging correction factor, is in the range 5--20\%, depending on the resonance mass \mx.
The \nsub selection efficiency uncertainty depends on how many \nsub tags are used, two for the fully-merged (26--30\% uncertainty) and one for the semi-resolved analysis (13--15\% uncertainty). This includes an additional uncertainty in the $\tau_{21}$ scale factor, determined using simulations, for jets with \pt higher than those in the \ttbar events used for the evaluation of this systematic.

Scale factors are used to correct the signal events yields so their \Hbbt and DeepCSV discriminator efficiencies are the same as for data. The \Hbbt and the DeepCSV discriminator scale factors are taken to be 100\% correlated. The associated uncertainty is 2--9\%~\cite{Sirunyan:2017ezt}, depending on the \Hbbt and requirement threshold and jet \pt, and is propagated to the total uncertainty in the signal yield.

The impact of the theoretical scale uncertainties and PDF uncertainties, the latter derived using the {PDF4LHC} procedure~\cite{Butterworth:2015oua} and the {NNPDF3.0} PDF sets, is estimated to be 0.1--3\%. These uncertainties affect the product of the signal acceptance and the selection efficiency. The scale and the PDF uncertainties have negligible impact on the signal \mjjjs distributions.
Additional systematic uncertainties associated with the pileup modelling (1--2\%, based on a 4.6\% variation on the $\Pp\Pp$ total inelastic cross section) and with the integrated luminosity determination (2.5\%)~\cite{CMS-PAS-LUM-17-001}, are applied to the signal yield.

The systematic uncertainty on the trijet invariant mass cut was calculated by comparing the cut efficiency for Pythia and Herwig bulk graviton samples, and is equivalent to a 0.5\% systematic.

The systematic uncertainty applied to the signal is also applied to the \ttjets background in the semi-resolved analysis, as appropriate. The total uncertainty in the \ttjets background is 11--15\%, of which 6\% derives from the uncertainty in the \ttjets cross section.

The main source of uncertainty for the multijet background is due to the statistical uncertainty in the fit to the $R_\text{p/f}$ ratio performed in the \PH jet mass sidebands. This uncertainty, amounting to 2--10\%, is fully correlated between all \mjjjs bins. Additional statistical uncertainties on the background shape and yield in the signal region result from the finite statistics of the multijets samples in the antitag region and are evaluated using the Barlow--Beeston Lite method~\cite{BarlowBeeston,BBLite}.
These uncertainties are small as compared with the uncertainty on the $R_\text{p/f}$ ratio, and are uncorrelated from bin to bin.

\section{Results\label{sec:Results}}

This analysis extends the search for a resonance $X$ decaying to $\PH\PH \to \bbbar\bbbar$ with two boosted \PH jets~\cite{CMS-B2G-16-026} to cover the semi-resolved topology involving one boosted \PH jet and one resolved \Hbb decay reconstructed using two $\cPqb$ jets. An $\PH\PH$ signal would appear as an excess of events over estimated background in the \mjjjs spectra of the different signal event categories, as discussed in Section~\ref{sec:SigModelBkgEst}.

\begin{figure}[thb!]
  \centering
    \includegraphics[width=0.48\textwidth]{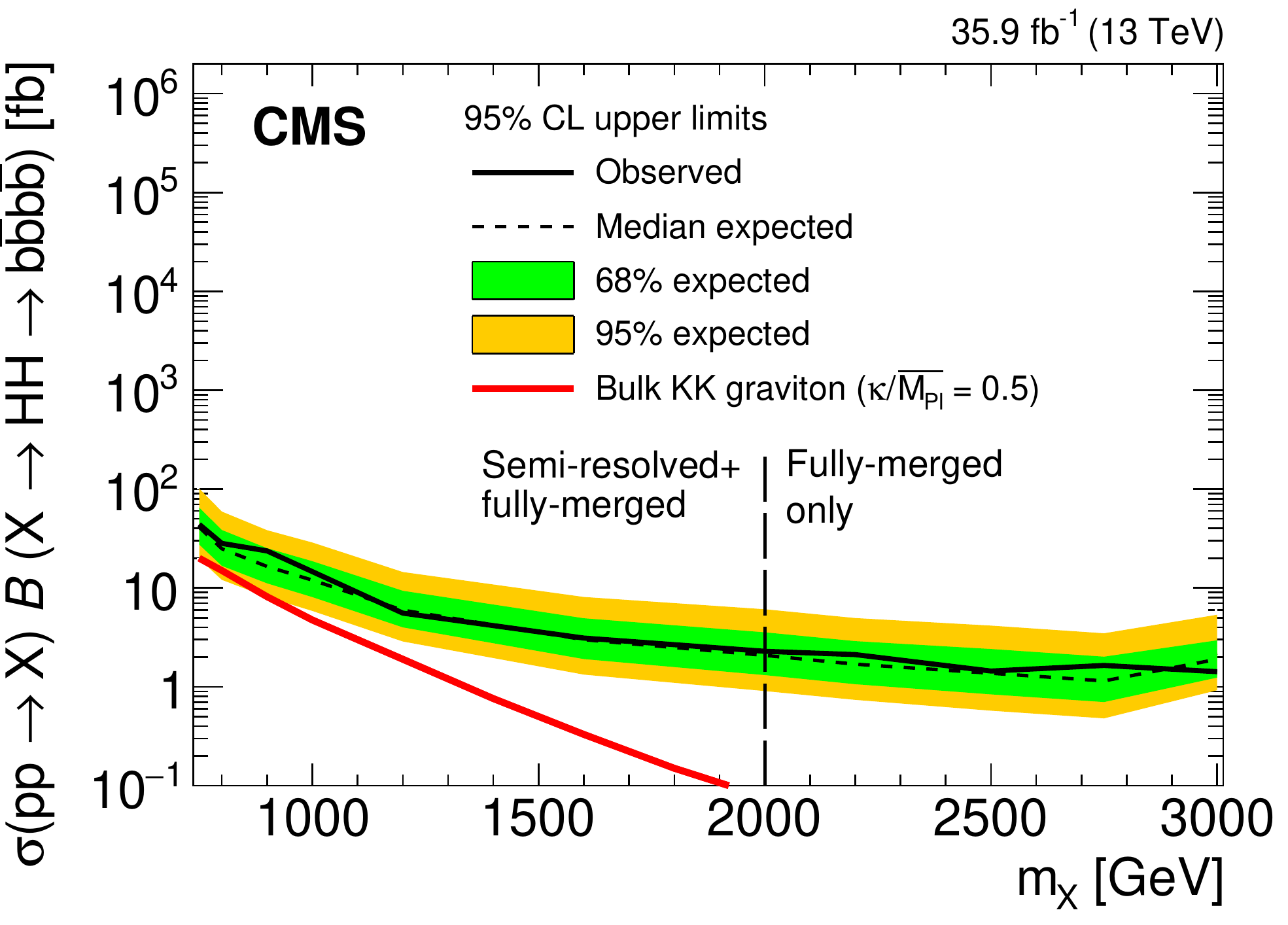}
    \includegraphics[width=0.48\textwidth]{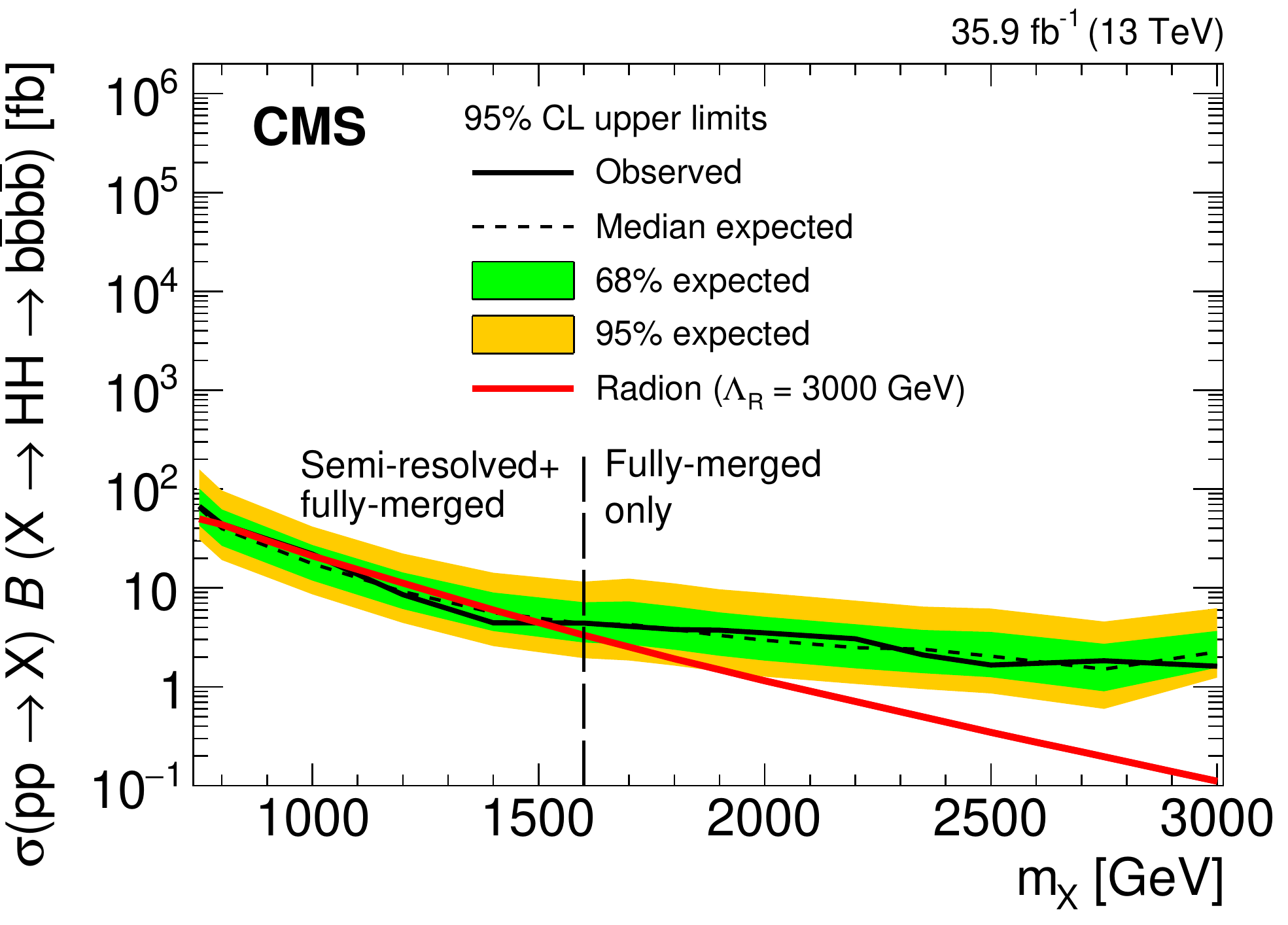}
  \caption{The upper limits for a bulk graviton (\cmsLeft) and radion (\cmsRight), combining the fully-merged and the semi-resolved analysis (where the events used in the fully-merged analysis are not considered in the semi-resolved analysis). The inner (green) and the outer (yellow) bands indicate the regions containing the 68 and 95\%, respectively, of the distribution of limits expected under the background-only hypothesis. The theoretical predictions are shown as the red lines. Results above 2000 (1600)\GeV for the bulk graviton (radion) are taken directly from the fully-merged analysis~\cite{CMS-B2G-16-026}.}
  \label{fig:1p12p1BG}
\end{figure}

\begin{figure}[thb!]
  \centering
    \includegraphics[width=0.65\textwidth]{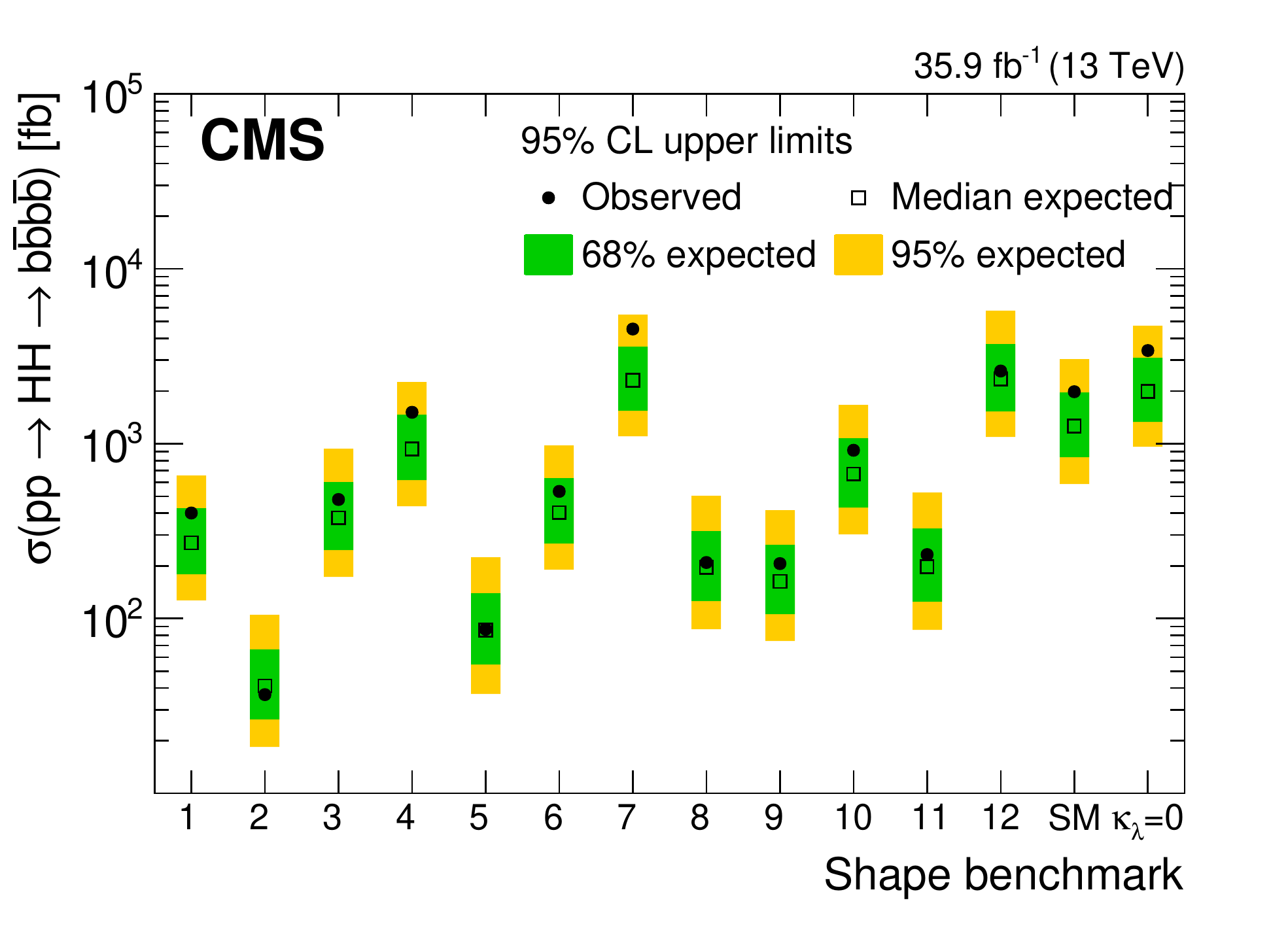}
  \caption{The observed and expected upper limits for non-resonant $\PH\PH$ production in the standard model, the model with $\kappa_{\lambda}=0$, and other shape benchmarks (1--12), combining the fully-merged selection and the semi-resolved selection (where the events used in the fully-merged analysis are not considered in the semi-resolved analysis). The inner (green) and the outer (yellow) bands indicate the regions
  containing the 68 and 95\%, respectively, of the distribution of limits
  expected under the background-only hypothesis.}
  \label{fig:1p12p1NR}
\end{figure}

The binned \mjjjs distributions of the signal and the background are fitted to the data, as shown in Fig.~\ref{fig:Alphabet_Rpf_semiresolved_rejectBoosted} (right), and examined for an excess of events above the predicted background. The data were found to be consistent with the expected background predictions.
Upper limits on the product of the signal cross sections and branching fractions are obtained using the profile likelihood as a test statistic~\cite{ATLASCMSComb}. The systematic uncertainties are treated as nuisance parameters and are profiled in the minimization of the negative of the logarithm of the profile likelihood ratio and the distributions of the likelihood ratio are calculated using the asymptotic approximation~\cite{AsympCLs} of the procedure reported in Refs.~\cite{CLS1,CLS2}.
Upper limits at 95\% confidence level are set on the product of the production cross section and the branching fractions $\sigma(\Pp\Pp\to\PX) \mathcal{B}(\PX\to\bbbar\bbbar)$.

Results are obtained using a statistical combination of the semi-resolved and fully-merged event categories for the bulk graviton having a mass between 750--2000\GeV, and a radion with a mass between 750--1600\GeV. Above these mass ranges, the inclusion of the semi-resolved events does not appreciably improve the search sensitivity, as evidenced from the expected limit values.
The limits on $\sigma(\Pp\Pp\to\PX) B(\PX\to\bbbar\bbbar)$ are shown in Fig.~\ref{fig:1p12p1BG}, and tabulated in Tables~\ref{tab:1p12p1BG} and~\ref{tab:1p12p1Rad} for the bulk graviton and the radion, respectively. The limits for $\mx > 2000\GeV$ for the bulk graviton, and $\mx > 1600\GeV$ for the radion are those from the fully-merged analysis of Ref.~\cite{CMS-B2G-16-026}.

For the interpretation of the results, this paper uses the scenario of Ref.~\cite{Fitzpatrick:2007qr} to describe a KK graviton, where the propagation of SM fields is allowed in the bulk, and follows the characteristics of the SM gauge group, with the right-handed top quark localized near the TeV brane. The radion is an additional element of WED models that is needed to stabilize the size of the extra dimension $l$.
The theoretical cross sections for $\sigma(\Pp\Pp\to\PX) B(\PX\to\bbbar\bbbar)$ are calculated using $\kappa/\overline{\Mpl}=0.5$ for the bulk gravitons and $\LambdaR=3\TeV$ for the radions, of different masses. For these values of $\kappa/\overline{\Mpl}$ and \LambdaR, the branching fractions $\mathcal{B}(\PX\to\bbbar\bbbar)$ are  10 and 23\%, for the graviton and the radion, respectively~\cite{Oliveira:2014kla}.
As shown in Fig.~\ref{fig:1p12p1BG} (right), a radion having a mass between 1000 and 1500\GeV is excluded at 95\% confidence level for $\LambdaR=3\TeV$.

The improvement in the upper limits on $\sigma(\Pp\Pp\to\PX) B(\PX \to \bbbar\bbbar)$ due to the inclusion of the semi-resolved event category between 18\% and 7\%, for a bulk graviton in the mass range 750--2000\GeV. A much larger improvement---between 55\% and 8\%---is seen for a radion in the mass range 750--1600\GeV. This can be attributed to the two pseudorapidity intervals, $\abs{\Delta\eta} < 1$ and $1 \le \abs{\Delta\eta} \le 2$, utilized in the semi-resolved event selection, with the lower pseudorapidity interval having a better signal to background ratio for a spin-0 radion, because of the angular distribution of its decay products.

\begin{table}[h]
  \centering
  \topcaption{The observed and expected upper limits on the products of the cross sections and branching fraction $\sigma(\Pp\Pp\to\PX) B(\PX \to \PH\PH \to \bbbar\bbbar)$ for a bulk graviton from the combination of the fully-merged and semi-resolved channels (where the events used in the fully-merged analysis are not considered in the semi-resolved analysis). Results above 2000\GeV are taken directly from the fully-merged analysis~\cite{CMS-B2G-16-026}.}
\cmsTable{
    \begin{tabular}{ccccccc}
      \hline
      Mass & Obs. lim. & Exp. lim. & +Exp (68\%) & -Exp (68\%) & +Exp (95\%) & -Exp (95\%) \\
      ({\GeVns}) & (\!\unit{fb}) & (\!\unit{fb}) & (\!\unit{fb}) & (\!\unit{fb})& (\!\unit{fb})& (\!\unit{fb}) \\ \hline
      750  & 43.9 & 41.0 & 27.4 & 64.8 & 19.6 & 101\\
      800  & 28.2 & 24.9 & 16.7 & 38.5 & 12.1 & 59.2\\
      900  & 23.6 & 16.4 & 11.1 & 25.2 & 8.1 & 38.4\\
      1000 & 14.6 & 11.9 & 8.0 & 18.6 & 5.9 & 28.8\\
      1200 & 5.5 & 5.9 & 3.9 & 9.3 & 2.9 & 14.5\\
      1600 & 3.1 & 3.0 & 1.9 & 4.9 & 1.3 & 8.1\\
      2000 & 2.2 & 2.0 & 1.3 & 3.5 & 0.9 & 6.1\\
      2500 &1.4 & 1.4 & 0.8 & 2.4 & 0.5 & 4.2\\
      3000 &1.4 & 1.9 & 1.2 & 3.0 & 0.9 & 5.3\\
      \hline
    \end{tabular}
}
\label{tab:1p12p1BG}
\end{table}

\begin{table}[h]
  \centering
  \topcaption{The observed and expected upper limits on the products of the cross sections and branching fraction $\sigma(\Pp\Pp\to\PX) B(\PX \to \PH\PH \to \bbbar\bbbar)$ for a radion from the combination of the fully-merged and semi-resolved channels (where the events used in the fully-merged analysis are not considered in the semi-resolved analysis). Results above 1600\GeV for the radion are taken directly from the fully-merged analysis~\cite{CMS-B2G-16-026}.}
\cmsTable{
    \begin{tabular}{ccccccc}
      \hline
      Mass &  Obs. lim. & Exp. lim. & +Exp (68\%) & -Exp (68\%) & +Exp (95\%) & -Exp (95\%) \\
      ({\GeVns}) & (\!\unit{fb}) & (\!\unit{fb}) & (\!\unit{fb}) & (\!\unit{fb})& (\!\unit{fb})& (\!\unit{fb}) \\ \hline
      750  & 67.0 & 64.5 & 42.8 & 101 & 30.9 & 158\\
      800  & 44.3 & 39.8 & 26.6 & 62.2 & 19.1 & 96.8\\
      900  & 31.2 & 28.6 & 19.8 & 43.1 & 15.9 & 65.1 \\
      1000 & 22.0 & 17.5 & 11.8 & 27.1 & 8.6 & 41.8\\
      1200 & 8.5 & 9.1 & 6.1 & 14.3 & 4.4 & 22.3\\
      1600 & 4.4 & 4.3 & 2.8 & 7.1 & 1.9 & 11.6\\
      2000 & 3.5 & 3.0 & 1.8 & 5.1 & 1.3 & 8.9\\
      2500 & 1.7 & 2.0 & 1.3 & 3.6 & 0.8 & 6.1\\
      3000 & 1.6 & 2.3 & 1.6 & 3.7 & 1.2 & 6.2\\
      \hline
    \end{tabular}
}
\label{tab:1p12p1Rad}
\end{table}

\begin{table}[h]
  \centering
  \topcaption{The observed and expected upper limits on the cross section $\sigma(\Pp\Pp \to \PH\PH \to \bbbar\bbbar)$ for the non-resonant shape benchmark models (1--12), the SM, and the $\kappa_{\lambda}=0$ $\PH\PH$ productions, combining fully-merged and semi-resolved channels (where the events used in the fully-merged analysis are not considered in the semi-resolved analysis).}\label{tab:2p1limcomb}
\cmsTable{
      \begin{tabular}{ccccccc}
        \hline
        Shape  &  Obs. lim. & Exp. lim. & +Exp (68\%) & -Exp (68\%) & +Exp (95\%) & -Exp (95\%) \\
        benchmark & (\!\unit{fb}) & (\!\unit{fb}) & (\!\unit{fb}) & (\!\unit{fb}) & (\!\unit{fb}) & (\!\unit{fb}) \\ \hline
        1 & 401 & 271 & 179 & 428 & 127 & 660\\
        2 & 36.7 & 41.0 & 26.5 & 66.3 & 18.5 & 105\\
        3 & 479 & 376 & 247 & 601 & 173 & 936\\
        4 & 1510 & 932 & 618 & 1460 & 438 & 2240\\
        5 & 86.6 & 85.9 & 54.4 & 140 & 37.0 & 225\\
        6 & 533 & 403 & 268 & 637 & 190 & 978\\
        7 & 4520 & 2300 & 1530 & 3580 & 1100 & 5470\\
        8 & 209 & 196 & 126 & 317 & 87.2 & 504\\
        9 & 206 & 163 & 106 & 264 & 74 & 415\\
        10 & 916 & 670 & 433 & 1070 & 302 & 1660\\
        11 & 232 & 198 & 125 & 326 & 85.9 & 526\\
        12 & 2600 & 2330 & 1530 & 3700 & 1090 & 5750\\
        SM & 1980 & 1260 & 833 & 1970 & 589 & 3030\\
        $\kappa_{\lambda}=0$ & 3404 & 1989 & 3092 & 1334 & 4732 & 960 \\
        \hline
      \end{tabular}
    }
\end{table}

In addition, both the fully-merged and the semi-resolved analyses look for non-resonant $\PH\PH$ production.
The observed and expected upper limits are presented in Table~\ref{tab:2p1limcomb} for the semi-resolved and the fully-merged signal categories combined, also depicted in Fig.~\ref{fig:1p12p1NR}.
The observed and expected limits are respectively, 179 and 114 times the product of the SM cross sections and branching fractions.
The new limits are better by about a factor of three for benchmark 2 and a factor of two for benchmark 5, with significant improvements for benchmarks 8, 9, and 11, compared to existing measurements~\cite{CMS-HIG-17-008,CMSHHbbtatau13TeV2016data}.
The increased sensitivity of this analysis to certain benchmarks is due to the higher level of destructive interference among the $\PH\PH$ production processes close to the kinematic threshold, which leads to a corresponding shift of the $\PH\PH$ mass spectrum towards higher values.
This leads to, on average, a higher \pt of the Higgs bosons, and hence in the sensitivity of this analysis, which identifies Higgs bosons using boosted techniques.

\section{Summary\label{sec:Summary}}

A search is presented for the pair production of standard model Higgs bosons ($\PH\PH$), both decaying to a bottom quark-antiquark pair ($\bbbar$), using data from proton-proton collisions at a centre-of-mass energy of 13\TeV and corresponding to an integrated luminosity of \intLumi.
The search is conducted in the region of phase space where at least one of the Higgs bosons has a large Lorentz boost, so that the $\PH\to\bbbar$ decay products are collimated to form a single jet, an \PH jet.
The search combines events with one $\PH$ jet plus two $\cPqb$ jets with events having two $\PH$ jets, thus adding sensitivity to the previous analysis~\cite{CMS-B2G-16-026}.

The results of the search are compared with predictions for the resonant production of a narrow Kaluza--Klein bulk graviton and a narrow radion in warped extradimensional models. The search is also sensitive to several beyond standard model non-resonant $\PH\PH$ production scenarios. Such cases may arise either when an off-shell massive resonance produced in proton-proton collisions decays to $\PH\PH$, or through beyond standard model effects in the Higgs boson coupling parameters. The results are interpreted in terms of upper limits on the product of the cross section for the respective signal processes and the branching fraction to $\PH\PH \to \bbbar\bbbar$, at 95\% confidence level.

The upper limits range from 43.9 to 1.4\unit{fb} for the bulk graviton and from 67 to 1.6\unit{fb} for the radion for the mass range 750--3000\GeV. Depending on the mass of the resonance, these limits improve upon the results of Ref.~\cite{CMS-B2G-16-026} by up to 18\% for the bulk graviton and up to 55\% for the radion.

The non-resonant production of Higgs boson pairs is modelled using an effective Lagrangian with five coupling parameters.
The upper limit corresponding to the standard model values of the coupling parameters is placed at 1980\unit{fb}, which is 179 times the prediction.
In addition, upper limits in the range of 4520 to 36.7\unit{fb} are set on twelve shape benchmarks, \ie representative sets of the five coupling parameters~\cite{Carvalho2016}.
These are the first limits on non-resonant Higgs boson pair-production signals using boosted topologies, and are the most stringent limits to date for the shape benchmarks 2, 5, 8, 9, and 11.

\begin{acknowledgments}
  We congratulate our colleagues in the CERN accelerator departments for the excellent performance of the LHC and thank the technical and administrative staffs at CERN and at other CMS institutes for their contributions to the success of the CMS effort. In addition, we gratefully acknowledge the computing centres and personnel of the Worldwide LHC Computing Grid for delivering so effectively the computing infrastructure essential to our analyses. Finally, we acknowledge the enduring support for the construction and operation of the LHC and the CMS detector provided by the following funding agencies: BMBWF and FWF (Austria); FNRS and FWO (Belgium); CNPq, CAPES, FAPERJ, FAPERGS, and FAPESP (Brazil); MES (Bulgaria); CERN; CAS, MoST, and NSFC (China); COLCIENCIAS (Colombia); MSES and CSF (Croatia); RPF (Cyprus); SENESCYT (Ecuador); MoER, ERC IUT, and ERDF (Estonia); Academy of Finland, MEC, and HIP (Finland); CEA and CNRS/IN2P3 (France); BMBF, DFG, and HGF (Germany); GSRT (Greece); NKFIA (Hungary); DAE and DST (India); IPM (Iran); SFI (Ireland); INFN (Italy); MSIP and NRF (Republic of Korea); MES (Latvia); LAS (Lithuania); MOE and UM (Malaysia); BUAP, CINVESTAV, CONACYT, LNS, SEP, and UASLP-FAI (Mexico); MOS (Montenegro); MBIE (New Zealand); PAEC (Pakistan); MSHE and NSC (Poland); FCT (Portugal); JINR (Dubna); MON, RosAtom, RAS, RFBR, and NRC KI (Russia); MESTD (Serbia); SEIDI, CPAN, PCTI, and FEDER (Spain); MOSTR (Sri Lanka); Swiss Funding Agencies (Switzerland); MST (Taipei); ThEPCenter, IPST, STAR, and NSTDA (Thailand); TUBITAK and TAEK (Turkey); NASU and SFFR (Ukraine); STFC (United Kingdom); DOE and NSF (USA).

  \hyphenation{Rachada-pisek} Individuals have received support from the Marie-Curie programme and the European Research Council and Horizon 2020 Grant, contract No. 675440 (European Union); the Leventis Foundation; the A. P. Sloan Foundation; the Alexander von Humboldt Foundation; the Belgian Federal Science Policy Office; the Fonds pour la Formation \`a la Recherche dans l'Industrie et dans l'Agriculture (FRIA-Belgium); the Agentschap voor Innovatie door Wetenschap en Technologie (IWT-Belgium); the F.R.S.-FNRS and FWO (Belgium) under the ``Excellence of Science - EOS" - be.h project n. 30820817; the Ministry of Education, Youth and Sports (MEYS) of the Czech Republic; the Lend\"ulet (``Momentum") Programme and the J\'anos Bolyai Research Scholarship of the Hungarian Academy of Sciences, the New National Excellence Program \'UNKP, the NKFIA research grants 123842, 123959, 124845, 124850 and 125105 (Hungary); the Council of Science and Industrial Research, India; the HOMING PLUS programme of the Foundation for Polish Science, cofinanced from European Union, Regional Development Fund, the Mobility Plus programme of the Ministry of Science and Higher Education, the National Science Center (Poland), contracts Harmonia 2014/14/M/ST2/00428, Opus 2014/13/B/ST2/02543, 2014/15/B/ST2/03998, and 2015/19/B/ST2/02861, Sonata-bis 2012/07/E/ST2/01406; the National Priorities Research Program by Qatar National Research Fund; the Programa Estatal de Fomento de la Investigaci{\'o}n Cient{\'i}fica y T{\'e}cnica de Excelencia Mar\'{\i}a de Maeztu, grant MDM-2015-0509 and the Programa Severo Ochoa del Principado de Asturias; the Thalis and Aristeia programmes cofinanced by EU-ESF and the Greek NSRF; the Rachadapisek Sompot Fund for Postdoctoral Fellowship, Chulalongkorn University and the Chulalongkorn Academic into Its 2nd Century Project Advancement Project (Thailand); the Welch Foundation, contract C-1845; and the Weston Havens Foundation (USA).
\end{acknowledgments}

\bibliography{auto_generated}
\cleardoublepage \appendix\section{The CMS Collaboration \label{app:collab}}\begin{sloppypar}\hyphenpenalty=5000\widowpenalty=500\clubpenalty=5000\vskip\cmsinstskip
\textbf{Yerevan Physics Institute, Yerevan, Armenia}\\*[0pt]
A.M.~Sirunyan, A.~Tumasyan
\vskip\cmsinstskip
\textbf{Institut f\"{u}r Hochenergiephysik, Wien, Austria}\\*[0pt]
W.~Adam, F.~Ambrogi, E.~Asilar, T.~Bergauer, J.~Brandstetter, M.~Dragicevic, J.~Er\"{o}, A.~Escalante~Del~Valle, M.~Flechl, R.~Fr\"{u}hwirth\cmsAuthorMark{1}, V.M.~Ghete, J.~Hrubec, M.~Jeitler\cmsAuthorMark{1}, N.~Krammer, I.~Kr\"{a}tschmer, D.~Liko, T.~Madlener, I.~Mikulec, N.~Rad, H.~Rohringer, J.~Schieck\cmsAuthorMark{1}, R.~Sch\"{o}fbeck, M.~Spanring, D.~Spitzbart, A.~Taurok, W.~Waltenberger, J.~Wittmann, C.-E.~Wulz\cmsAuthorMark{1}, M.~Zarucki
% ...
\vskip\cmsinstskip
% ...
1:  Also at Vienna University of Technology, Vienna, Austria\\
% ...
\end{sloppypar}
\end{document}